\documentclass[conference]{IEEEtran}

\usepackage[utf8]{inputenc}
\usepackage[T1]{fontenc}
\usepackage{cite}
\usepackage{amsmath,amssymb,amsfonts}
\usepackage{upgreek}  
\usepackage{graphicx}
\usepackage{xspace}
\usepackage{subcaption}
\usepackage{booktabs}
\usepackage{url}
\usepackage{xcolor}
\usepackage{hyperref}
\def\BibTeX{{\rm B\kern-.05em{\sc i\kern-.025em b}\kern-.08em
    T\kern-.1667em\lower.7ex\hbox{E}\kern-.125emX}}

\newif\ifdoubleblind 
\doubleblindtrue 

\usepackage{pifont}
\newcommand{\good}{\textcolor{green!50!black}{\ding{51}}\xspace}
\newcommand{\bad}{\textcolor{red!70!black}{\ding{55}}\xspace}
\newcommand{\mixed}{\textcolor{orange!90!black}{$\boldsymbol{\sim}$}\xspace}

\usepackage{todonotes}

\newcommand{\luan}[1]{{\color{violet} \textbf{Luan:} #1}}
\newcommand{\added}[1]{{\color{green!50!black}#1}}

\newcommand{\francesco}[1]{{\color{gray} \textbf{Francesco:} #1}}
\newcommand{\intra}{\textit{intra}\xspace}
\newcommand{\inter}{\textit{inter}\xspace}
\newcommand{\pld}{\textit{PLD}\xspace}
\newcommand{\sbd}{\textit{SBD}\xspace}
\newcommand{\ddtw}{\textit{DDTW}\xspace}
\newcommand{\sage}{SAGE\xspace}
\newcommand{\sagetwo}{SAGE-2\xspace}
\newcommand{\kshape}{K-shape\xspace}
\newcommand{\kmeans}{K-means\xspace}

\newcommand{\kmeansdtw}{K-means(DTW)\xspace}
\newcommand{\random}{Random\xspace}

\newcommand{\tsfresh}{Tsfresh\xspace}
\ifdoubleblind
\newcommand{\plafrim}{\textcolor{blue}{System A}\xspace}
\newcommand{\polaris}{\textcolor{blue}{System B}\xspace}

\else
\newcommand{\plafrim}{PlaFRIM\xspace}

\newcommand{\polaris}{Polaris\xspace}

\fi

\newcommand{\numberofdatasets}{2\xspace}
\newcommand{\totalnumberoftraces}{$819{,}771$\xspace} 
\newcommand{\howmuchfaster}{on average $54\times$\xspace}
\newcommand{\numberofpublishedtraces}{$108{,}433$\xspace}
\newcommand{\improvementonstretch}{decrease I/O stretch by up to $24\%$ compared to the state of the art (round-robin placement)\xspace}
\newcommand{\clairvoyant}{\texttt{Clairvoyant}\xspace}
\newcommand{\fig}{Fig.\xspace}

\newcommand{\intragainsagetwovskshapewrite}{$9.1\%$\xspace}

\usepackage[most]{tcolorbox}
\newtcolorbox{rqbox}[1]{enhanced, breakable, colback=blue!5, colframe=blue!55!black,
  fonttitle=\bfseries, title={#1}, left=1mm, right=1mm, top=1mm, bottom=1mm}

\usepackage{tikz}
\usetikzlibrary{arrows.meta,positioning,fit,backgrounds,calc}
\tikzset{
  wfbox/.style={draw=black!55, rounded corners=2pt, fill=white, inner sep=2.5pt,
                align=center, font=\scriptsize},
  wfstep/.style={wfbox, fill=blue!4},
  wfout/.style={wfbox, fill=orange!10, draw=orange!60!black},
  wfarr/.style={-{Latex[length=4pt,width=3pt]}, draw=black!65, line width=0.5pt},
  wfpanel/.style={draw=black!30, dashed, rounded corners=3pt, inner sep=4pt},
  wftitle/.style={font=\scriptsize\bfseries, align=center},
}
\newcommand{\wfplot}[3][black!70]{\tikz[baseline=0pt,x=#2,y=1pt]{%
  \draw[#1,line width=0.55pt] plot coordinates{#3};}}
\newcommand{\wfcutplot}[3][black!70]{\tikz[baseline=0pt,x=#2,y=1pt]{%
  \foreach \x in {4,8}{\draw[black!35,dashed,line width=0.3pt] (\x,-1)--(\x,9);}
  \draw[#1,line width=0.55pt] plot coordinates{#3};}}
\newcommand{\wfdotplot}[3][black!70]{\tikz[baseline=0pt,x=#2,y=1pt]{%
  \draw[#1,line width=0.55pt] plot[mark=*,mark size=0.7pt] coordinates{#3};}}
\newcommand{\wfbars}[3][black!70]{\tikz[baseline=0pt,x=#2,y=4.5pt]{%
  \draw[black!35,line width=0.3pt] (-0.6,0)--(3.6,0);
  \foreach \i/\v in {#3}{\draw[#1,line width=2.2pt] (\i,0)--(\i,\v);}}}
\newcommand{\tgen}[1]{{\ttfamily\textup{#1}}}
\newcommand{\wtI}{(0,1)(1,5)(2,8)(3,3)(4,1)(5,4)(6,7)(7,2)(8,0)(9,3)(10,6)(11,2)(12,0)}
\newcommand{\wtII}{(0,0)(1,1)(2,7)(3,7)(4,1)(5,0)(6,1)(7,6)(8,7)(9,1)(10,0)(11,1)(12,0)}
\newcommand{\wtIII}{(0,2)(1,3)(2,2)(3,4)(4,3)(5,2)(6,8)(7,7)(8,2)(9,3)(10,2)(11,3)(12,2)}
\newcommand{\wtIV}{(0,3)(1,5)(2,7)(3,5)(4,3)(5,1)(6,0)(7,1)(8,3)(9,5)(10,8)(11,5)(12,3)}

\usepackage{comment}
\newcommand{\rrtext}[1]{\ifx\researchreport\undefined\else#1\fi}
\ifx\researchreport\undefined
  \excludecomment{rronly}
  \includecomment{cponly}
\else
  \includecomment{rronly}
  \excludecomment{cponly}
\fi

\makeatletter
\newcommand{\linebreakand}{%
  \end{@IEEEauthorhalign}
  \hfill\mbox{}\par
  \mbox{}\hfill\begin{@IEEEauthorhalign}
}
\makeatother

\begin{document}

\title{Fast and Actionable Classification of HPC I/O Time Series}


\author{\IEEEauthorblockN{Maxime Gonthier}
  \IEEEauthorblockA{\textit{Inria}\\
  France \\
  maxime.gonthier@inria.fr}
  \and
  \IEEEauthorblockN{Luan Teylo}
  \IEEEauthorblockA{\textit{Inria}\\
  France \\
  luan.teylo@inria.fr}
  \and
  \IEEEauthorblockN{Francesco Antici}
  \IEEEauthorblockA{\textit{RIKEN Center for Computational Science}\\
  Japan \\
  francesco.antici@riken.jp}
  \linebreakand
  \IEEEauthorblockN{Mihail Popov}
  \IEEEauthorblockA{\textit{Inria}\\
  France \\
  mihail.popov@inria.fr}
  \and
  \IEEEauthorblockN{Francieli Boito}
  \IEEEauthorblockA{\textit{University of Bordeaux, Inria}\\
  France \\
  francieli.zanon-boito@u-bordeaux.fr}}

\maketitle

\begin{abstract} 
Input/output (I/O) is a major bottleneck for high-performance computing (HPC) performance.
One of the challenges is the contention between jobs when sharing the parallel file system. However, tackling this issue depends on knowing the temporal behavior of jobs. In this paper, we propose a clustering-based approach to characterize jobs' temporal I/O behavior. 
We propose \sage, a lightweight method that cuts I/O traces into short windows and learns an alphabet from them. The trace then becomes a sequence of symbols, which we call ``genome'', and we use clustering to separate them into classes. 
%
We evaluate \sage and its two-stage variant \sagetwo on \totalnumberoftraces real traces from \numberofdatasets HPC systems. 
%
We compare them against other techniques from the state of the art 
on three metrics, including one we propose. 
Compared to \kshape, the best baseline available, \sagetwo lowers the median \intra-cluster metrics by \intragainsagetwovskshapewrite on average, and achieves this in $2$ minutes of clustering time versus $3.8$ hours, a $109\times$ speedup.
More importantly, clusters created by \sage and \sagetwo carry enough information to act on: we evaluate the use of the cluster centroids for I/O scheduling, and show that our approach can \improvementonstretch.
%

\end{abstract}



\section{Introduction}\label{sec:introduction}

High-performance computing (HPC) systems typically rely on a parallel file system (PFS) (such as BeeGFS~\cite{beegfs} and 
Lustre~\cite{lustre}) to enable access to persistent data from system resources. 
The PFS is shared among all the running jobs, and concurrent I/O can create \emph{interference}~\cite{7516071,Trochon2026OnTI,iosets}, which degrades application performance.
Depending on their characteristics, some applications 
impose a heavier slowdown on the others, and 
executing certain operations together may even lead to degradation of overall system performance~\cite{227665,9139855,Trochon2026OnTI}.

Measurements on a production supercomputer show this effect~\cite{A_Deep_Look_into_the_Temporal_IO_Behavior}: as I/O concurrency increases, contention causes system performance to decline.
Interference leads to a waste of expensive, power-hungry compute resources, which are often idle while waiting for I/O to finish, 
especially when the system performance is impacted. Moreover, interference causes high performance variability~\cite{9229642,9355272,8752753}, which complicates
resource management and performance explainability, as application performance depends not only on its operations, but also on the behavior of other applications.

I/O scheduling approaches have been proposed to alleviate this problem~\cite{iosets,10.1145/3581784.3607041,10.1109/IPDPS.2015.116,7307592,7877148}. They explore techniques such as delaying job start times~\cite{10.1109/IPDPS.2015.116,7877148},
preventing concurrent I/O activity of specific jobs~\cite{iosets,10.1145/3581784.3607041,7307592} and assigning priorities~\cite{iosets}. However, for those techniques to work,
they require the knowledge (or, rather, the \emph{prediction}) of \emph{when} jobs issue I/O requests. In this paper, we work to provide this information \emph{without} depending on assumptions about job behavior, such as periodicity~\cite{10579160,10.1109/IPDPS.2015.116}.


We argue against trying to predict the exact temporal behavior for all jobs, but rather to \emph{classify} them. 
Indeed, 
the exact temporal I/O behavior of a job is difficult to predict and likely to change across executions, as it is impacted by performance variability 
and I/O interference.
Moreover,
%
\ifdoubleblind
related
\fi
 work~\cite{A_Deep_Look_into_the_Temporal_IO_Behavior} has indicated that a relatively small number of patterns can represent most HPC jobs. 
A scheduling algorithm could use the general behavior of such classes as proxies for job behavior in order to mitigate interference. 
Finally, such a classification
would add great value to the community, allowing for studies of the temporal behavior of HPC applications. 
Such a classification also allows for properly representing a system's workload when creating benchmarks, for example, allowing for more realistic evaluation of new proposals.

In this paper, we extensively study the problem of 
clustering\footnote{We use the terms \emph{cluster} and \emph{class} interchangeably throughout the paper.} I/O behavior  
and present a use case of I/O scheduling that uses our classification to improve I/O performance 
by mitigating interference. Our main contributions are as follows.

\begin{itemize}
    \item  We propose two techniques, namely Shape-Amplitude Genome (\sage) and \sagetwo, which rely on a clustering-based 
    encoding of I/O time series that learns an alphabet from short windows, 
    categorizing them by shape and amplitude. 
    
    \item We introduce Phase-Level Distance (\pld),
    a noise-resistant metric that captures both amplitude and temporal shape differences. 

    \item Our extensive evaluation includes approaches from the state of the art, namely \kmeans, \kshape, \tsfresh, and \kmeansdtw as well as multiple metrics. It is based on real data sets from \numberofdatasets systems, comprising
    \totalnumberoftraces read and write traces, and shows that \sage and \sagetwo improve clustering quality results over the state of the art, while also being \howmuchfaster faster. 

    \item We show the usefulness of our classification by evaluating an I/O scheduling algorithm that takes the classes' behaviors into 
    consideration for placing jobs 
    on distributed PFS resources. 
    Our results show that we \improvementonstretch.

\ifdoubleblind
    \item We create and publicly release a data set containing the I/O traces of \numberofpublishedtraces HPC jobs in order to support further work in this field.
\fi         
\end{itemize}

The paper is organized as follows.
Section~\ref{sec:related_work} discusses related work.
We present \sage and \sagetwo in Section~\ref{sec:sage}.
Section~\ref{sec:baseline_and_workload} describes the baselines and the data sets used in our experiments.
Section~\ref{sec:evaluation} evaluates the quality of the resulting clusters.
Section~\ref{sec:simulation} uses these clusters to co-schedule jobs in a simulated system.

\section{Motivation} \label{sec:motivation}

In this paper we focus on classifying the I/O operations of production jobs to perform informed job scheduling and prevent interference. To achieve so, it is fundamental to thoroughly characterize job executions (e.g. phases, minimum, peaks, idle). This can be achieved by leveraging time series of I/O activity collected periodically during job execution. 
However, characterizing and clustering such data comes with notorious difficulties.

\begin{figure}[t]
  \centering
  \begin{minipage}[t]{0.482\columnwidth}
    \vspace{0pt}
    \includegraphics[width=\linewidth]{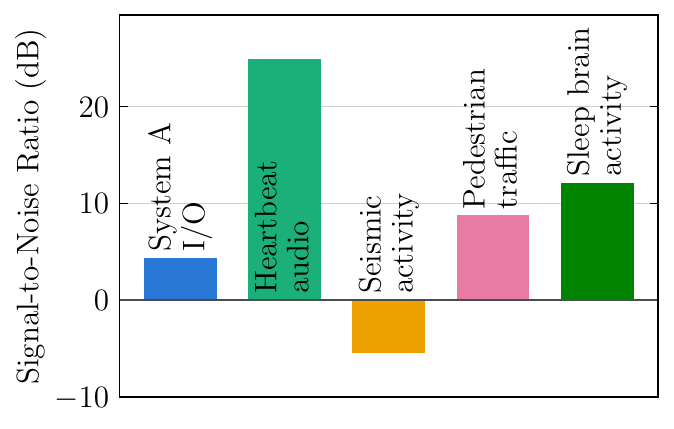}
  \end{minipage}
  \hfill
  \begin{minipage}[t]{0.498\columnwidth}
    \vspace{0pt}
    \includegraphics[width=\linewidth]{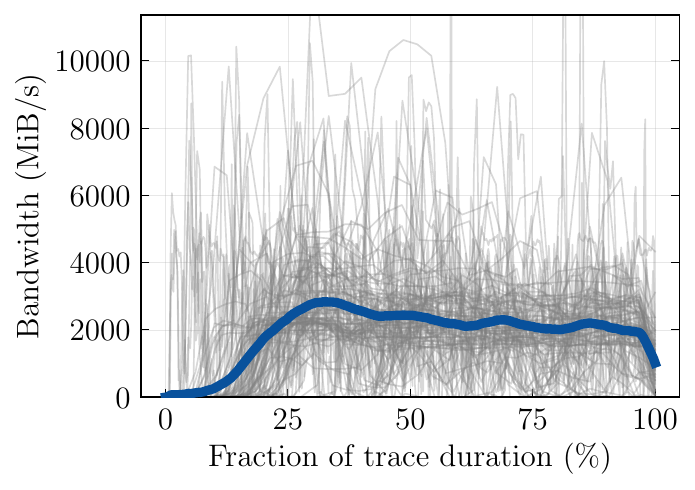}
  \end{minipage}
  \par
  \begin{minipage}[t]{0.482\columnwidth}
    \subcaption{Signal-to-noise ratio of the \plafrim\footnote{Machine names in blue were omitted for double-blind review.}
 I/O time series and of four data sets commonly used to evaluate clustering algorithms.}
    \label{fig:snr}
  \end{minipage}
  \hfill
  \begin{minipage}[t]{0.498\columnwidth}
    \subcaption{One of the clusters created by \kmeans for the \plafrim data set. The centroid is shown in blue, and the gray lines show a random selection of 100 jobs that belong to this cluster.}
    \label{fig:bad_clustering_motivation}
  \end{minipage}
  \caption{I/O traces are noisy, and typical clustering techniques group them poorly.}
  \label{fig:motivation-snr-kmeans}
\end{figure}

\paragraph{Clustering I/O time series is challenging}
There are many techniques to cluster time series, but they rely on a certain quality of the signal in order to properly discriminate different behaviors~\cite{ts_clustering_survey}. 
I/O bandwidth over time has a lot of zeroes, values span multiple orders of magnitude~\cite{9355272,8752753}, and tend to be bursty~\cite{9820616}.
Moreover, job duration varies immensely, creating time series of different lengths~\cite{A_Deep_Look_into_the_Temporal_IO_Behavior}.
\fig~\ref{fig:snr} compares the signal-to-noise ratio of our I/O traces with that of data sets commonly used to evaluate clustering\footnote{Pulled from~\cite{ts_datasets}: Heartbeat audio~\cite{heartbeats_db_source}, Seismic activity~\cite{earthquakes_db_source}, Pedestrian traffic~\cite{pedestrian_db_source}, Sleep brain activity~\cite{sleep_db_source, sleep_db_site_source}.}.
\begin{rronly}
We smooth each trace with a centered moving average over 5 points and treat what the smoothing removes as noise. \francesco{Write a short sentence to explain why you do this and what you want to find.}
The mean is excluded, so a steady rate with jitter scores low.
The SNR is $10\log_{10}(\sigma^2_\text{smooth}/\sigma^2_\text{noise})$ in dB, and we report its median over the traces of each data set.
\end{rronly}
We can observe that I/O time series have the second worst signal-to-noise ratio.
It sits just above seismic activity, which is known to be very noisy. 
This makes I/O traces\footnote{Throughout this paper, we use ``traces'' or ``jobs'' to refer to I/O time series.} difficult to analyze and extract patterns from. 

\paragraph{Typical techniques fall short}
\fig~\ref{fig:bad_clustering_motivation} shows one of the clusters  created by \kmeans for the \plafrim data set. We can see the jobs were placed together because of their similar average performance, with their lines presenting very different shapes. That means this classification is \emph{not} useful for knowing when I/O will happen.
In this same classification, \kmeans created 11 clusters, but assigned 64\% of the jobs to just two of them.
Section~\ref{sec:evaluation} provides further examples of how existing techniques fail for I/O time series.

\begin{figure}[t]
  \centering
  \includegraphics[width=0.8\columnwidth]{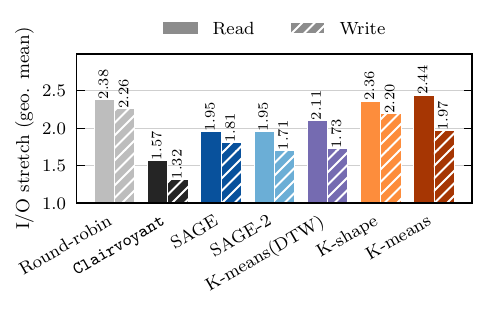}
  \caption{Results of an I/O scheduling algorithm using job clustering.
  We plot the geometric mean I/O stretch, i.e., the slowdown experienced by jobs due to sharing I/O resources. We compare the round-robin placement baseline, different clustering techniques, and a fictitious \clairvoyant solution that knows the actual I/O time series of every job.
  Workloads of 80 jobs on 4 object storage targets (OSTs) with a storage load $\rho = 1$ (see Section~\ref{sec:sim-workloads}). Solid bars use the read traces from \plafrim, and hatched bars the write traces. We later show that the clustering of \kmeansdtw is on average $63\times$ \emph{slower} than that of \sage (Table~\ref{tab:time}).}
  \label{fig:motivation-simulation}
\end{figure}

\paragraph{Classifying temporal I/O behavior is useful}
\fig~\ref{fig:motivation-simulation} previews results obtained with an I/O scheduling algorithm that uses jobs' clusters to avoid interference, 
discussed in Section~\ref{sec:use_case_scheduling}. The algorithm places jobs on PFS resources, trying not to place together jobs that will do I/O at the same time, and using cluster centroids as proxies for job behavior.
The plot shows the geometric mean I/O stretch, i.e., the slowdown experienced by a job when sharing the system with other jobs. We compare round-robin placement, \clairvoyant (which knows the jobs' exact time series), and results obtained using clusters generated by different techniques.
We can see that, by taking temporal behavior into consideration, \sage and \sagetwo decrease I/O stretch by up to 24\%
compared to round-robin placement.

\section{Related Work}\label{sec:related_work}

Related work tackles the I/O performance improvement by characterizing jobs' I/O behavior~\cite{iosets,10.1145/3581784.3607041,10.1109/IPDPS.2015.116,7307592,7877148}. 
%
Tarraf et al.~\cite{10579160} characterize periodic I/O using frequency techniques.
Omnisc'IO~\cite{7289462} builds a grammar of function called by the code, and uses it to predict future I/O activity.
A popular strategy 
is to apply supervised machine learning techniques~\cite{kim2023design,mckenna2016ml,IOscout,metis,Evalix,antici2026perf}. Unsupervised clustering is also used to classify jobs by energy consumption~\cite{ahmadzada:hal-05658205}, performance~\cite{antici2026perf}, failures~\cite{daradkeh2024hpc}, and combinations of multiple metrics~\cite{Halawa_2020}. 

Bang et al.~\cite{10.1145/3391812.3396270} combine logs from different layers of the I/O stack and extract summary features before clustering the jobs using \kmeans. 
AIOT~\cite{9820667} extracts I/O phases from traces using Discrete Wavelet Transform and clusters jobs 
based on metrics obtained from their I/O phases, such as peak bandwidth. 
In contrast, our technique keeps the time series format, instead of summarizing it by metrics, because we are interested in characterizing the \emph{temporal} I/O behavior. 

Betke and Kunkel~\cite{betke2021classifying} cluster one million jobs by encoding time series of I/O statistics 
into features and applying the Levenshtein distance as a similarity metric. They use more information than we do, encoding per-node data and metadata activity into a single time series. 
 In contrast, we classify one metric at a time. \emph{The idea is that a job would belong to different classes regarding its read/write/metadata activity over time.} Our classification thus provides information that is readily exploitable by optimization techniques. As discussed, we focus on being able to tell \emph{when} I/O happens.

 To process the large number of traces, Betke and Kunkel i) randomly select a small subset for clustering, ii) train a classification model to predict classes of jobs, and iii) use the model to classify the rest of the data set. They also compress the traces to the granularity of tens of minutes. We were not faced with this issue, and one of the reasons is that \sage is faster than most other techniques, as shown in Section~\ref{sec:evaluation}. Moreover, our clustering-based encoding achieves compression without sacrificing details.

Boito et al.~\cite{A_Deep_Look_into_the_Temporal_IO_Behavior} use K-means (with Dynamic Time Warping (DTW), as a distance metric) for clustering I/O time series to identify common behaviors. They separate the traces by length and peak performance, apply clustering on each group separately, and then merge the identified clusters by hand. In this paper, we propose an automatic approach, with no manual input.
Another important difference is that their classification focuses more on the \emph{shape}, as they merge classes from different amplitudes, while we work to keep both shape and \emph{amplitude}, as they are both important for I/O scheduling.

Once workloads are characterized, a common practice is to infer a job's I/O profile (e.g., its cluster) from the history of the user/group/project~\cite{9820667,user_based_io_profiling,mckenna2016ml,IOscout,Evalix}. Yazdani et al.~\cite{user_based_io_profiling} propose an I/O profiling tool based on past user activity and user-provided information. AIOT~\cite{9820667} builds a sequence of clusters by the same user, project, and scale, and uses a neural network to predict the next cluster. There is evidence in the literature for I/O behavior being somewhat stable for the same user/group/project~\cite{9820667,A_Deep_Look_into_the_Temporal_IO_Behavior,9910076}, and we can reasonably assume that predicting a class will be easier than predicting an exact time series, evidencing the importance of our contribution in this paper.
Predicting classes for jobs will be the focus of our future work.

\section{Translating Traces into ``Genomes'' and Clustering Them with \sage} \label{sec:sage}

The goal of Shape-Amplitude Genome (\sage) is to group jobs according to both \emph{when} they perform I/O during their execution and \emph{what their I/O profile looks like}. This will, for example, help a scheduler avoid placing jobs that perform I/O at the same time.
As shown in Section~\ref{sec:motivation}, this is challenging because I/O traces are noisy and their bandwidth spans several orders of magnitude. The key intuition behind \sage is that, although each trace is unique as a whole, its short segments are not. Similar patterns, such as brief I/O bursts, recur both within and across traces.
\sage learns these patterns once from windows extracted from all traces and assigns each pattern a symbol, much like a genome is represented using a small alphabet.
Each trace is then translated into a sequence of these symbols.
Replacing each window with a symbol also filters out much of the noise and produces a representation that is considerably shorter than the original trace, making subsequent clustering faster.

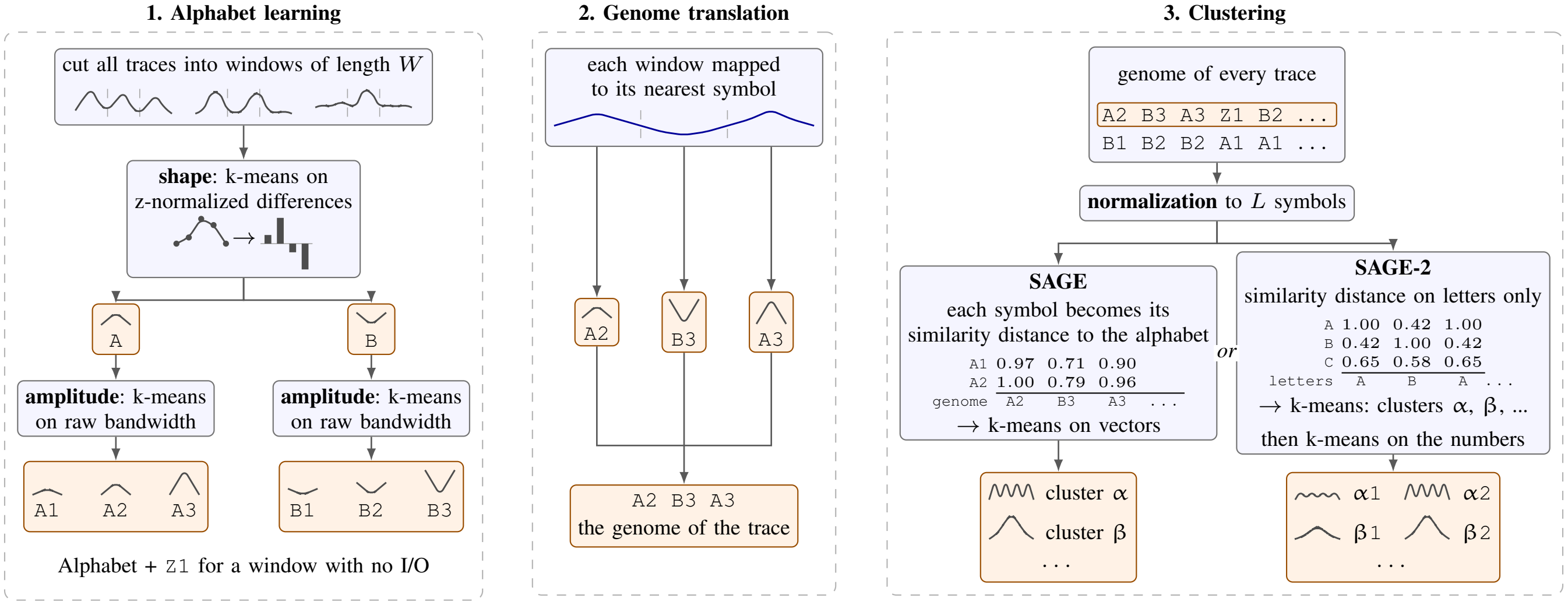
\begin{figure*}[t]
  \centering
  \begin{tikzpicture}[font=\scriptsize]

    \node[wfstep] (traces) at (2.9,0.57) {%
      cut all traces into windows of length $W$\\[3pt]
      \wfcutplot{2.6pt}{\wtI}\quad\wfcutplot{2.6pt}{\wtII}\quad\wfcutplot{2.6pt}{\wtIII}};

    \node[wfstep] (shapekm) at (2.9,-0.93) {%
      \textbf{shape}: k-means on\\ z-normalized differences\\[2pt]
      \begin{tabular}{@{}c@{\,}c@{\,}c@{}}
        \wfdotplot{4pt}{(0,0)(1,2)(2,8)(3,6)(4,0)} & $\rightarrow$ &
        \wfbars{4pt}{0/0.62,1/1.85,2/-0.62,3/-1.85} \\[-1pt]
      \end{tabular}};

    \node[wfout] (letterA) at (1.45,-2.18) {%
      \wfplot{2.4pt}{(0,0)(1,2)(2,4)(3,2)(4,0)}\\[-1pt] \tgen{A}};
    \node[wfout] (letterB) at (4.35,-2.18) {%
      \wfplot{2.4pt}{(0,4)(1,2)(2,0)(3,2)(4,4)}\\[-1pt] \tgen{B}};

    \node[wfstep] (ampA) at (1.45,-3.08) {%
      \textbf{amplitude}: k-means\\ on raw bandwidth};
    \node[wfstep] (ampB) at (4.35,-3.08) {%
      \textbf{amplitude}: k-means\\ on raw bandwidth};

    \node[wfout] (symA) at (1.45,-4.08) {%
      \begin{tabular}{@{}ccc@{}}
        \wfplot{2.4pt}{(0,0)(1,1)(2,2)(3,1)(4,0)} &
        \wfplot{2.4pt}{(0,0)(1,2)(2,4)(3,2)(4,0)} &
        \wfplot{2.4pt}{(0,0)(1,4)(2,8)(3,4)(4,0)} \\[-1pt]
        \tgen{A1} & \tgen{A2} & \tgen{A3} \\
      \end{tabular}};
    \node[wfout] (symB) at (4.35,-4.08) {%
      \begin{tabular}{@{}ccc@{}}
        \wfplot{2.4pt}{(0,2)(1,1)(2,0)(3,1)(4,2)} &
        \wfplot{2.4pt}{(0,4)(1,2)(2,0)(3,2)(4,4)} &
        \wfplot{2.4pt}{(0,8)(1,4)(2,0)(3,4)(4,8)} \\[-1pt]
        \tgen{B1} & \tgen{B2} & \tgen{B3} \\
      \end{tabular}};

    \node[font=\scriptsize, align=center] (znote) at (2.9,-4.88) {%
      Alphabet + \tgen{Z1} for a window with no I/O};

    \draw[wfarr] (traces) -- (shapekm);
    \draw[wfarr] (shapekm.south) -- ++(0,-0.25) -| (letterA.north);
    \draw[wfarr] (shapekm.south) -- ++(0,-0.25) -| (letterB.north);
    \draw[wfarr] (letterA) -- (ampA);
    \draw[wfarr] (letterB) -- (ampB);
    \draw[wfarr] (ampA) -- (symA);
    \draw[wfarr] (ampB) -- (symB);

    \node[wfstep] (onetrace) at (7.9,0.45) {%
      each window mapped\\to its nearest symbol\\[2pt]
      \wfcutplot[blue!60!black]{7pt}{\wtIV}};

    \node[wfout, inner sep=2pt] (w1) at (6.91,-2.1) {%
      \wfplot{2.4pt}{(0,0)(1,2)(2,4)(3,2)(4,0)}\\[-1pt] \tgen{A2}};
    \node[wfout, inner sep=2pt] (w2) at (7.9,-2.1) {%
      \wfplot{2.4pt}{(0,8)(1,4)(2,0)(3,4)(4,8)}\\[-1pt] \tgen{B3}};
    \node[wfout, inner sep=2pt] (w3) at (8.89,-2.1) {%
      \wfplot{2.4pt}{(0,0)(1,4)(2,8)(3,4)(4,0)}\\[-1pt] \tgen{A3}};

    \node[wfout] (genome) at (7.9,-4.3) {%
      \tgen{A2 B3 A3}\\[1pt]
      the genome of the trace};

    \draw[wfarr] ([xshift=-0.99cm]onetrace.south) -- (w1.north);
    \draw[wfarr] (onetrace.south) -- (w2.north);
    \draw[wfarr] ([xshift=0.99cm]onetrace.south) -- (w3.north);
    \draw[draw=black!65, line width=0.5pt] (w1.south) |- (7.9,-3.5);
    \draw[draw=black!65, line width=0.5pt] (w3.south) |- (7.9,-3.5);
    \draw[draw=black!65, line width=0.5pt] (w2.south) -- (7.9,-3.5);
    \draw[wfarr] (7.9,-3.5) -- (genome.north);

    \node[font=\scriptsize] (glab) at (13.95,0.7) {genome of every trace};
    \node[font=\scriptsize\ttfamily, draw=orange!60!black, fill=orange!10,
          rounded corners=1pt, inner sep=1.5pt, below=2pt of glab] (g1)
      {\tgen{A2 B3 A3 Z1 B2 $\dots$}};
    \node[font=\scriptsize\ttfamily, inner sep=1.5pt, below=1pt of g1] (g2)
      {\tgen{B1 B2 B2 A1 A1 $\dots$}};
    \begin{scope}[on background layer]
      \node[wfstep, fit=(g1)(g2)(glab)] (genomes) {};
    \end{scope}

    \node[wfstep] (resample) at (13.95,-0.75) {%
      \textbf{normalization} to $L$ symbols};

    \node[wfstep] (sage) at (12.15,-2.45) {%
      \textbf{\sage}\\[1pt] each symbol becomes its\\ similarity distance to the alphabet\\[3pt]
      {\tiny\setlength{\tabcolsep}{2pt}%
      \begin{tabular}{@{}r@{\ \,}cccc@{\,}c@{}}
        \tgen{A1} & $0.97$ & $0.71$ & $0.90$ & \\
        \tgen{A2} & $1.00$ & $0.79$ & $0.96$ & \\
        \cline{2-5}
        \tgen{genome} & \tgen{A2} & \tgen{B3} & \tgen{A3} & \tgen{$\dots$} \\
      \end{tabular}}\\[2pt]
      $\rightarrow$ k-means on vectors};
    \node[wfstep] (sagetwo) at (15.95,-2.45) {%
      \textbf{\sagetwo}\\[1pt] similarity distance on letters only\\[3pt]
      {\tiny\setlength{\tabcolsep}{2pt}%
      \begin{tabular}{@{}r@{\ \,}ccc@{\,}c@{}}
        \tgen{A} & $1.00$ & $0.42$ & $1.00$ & \\
        \tgen{B} & $0.42$ & $1.00$ & $0.42$ & \\
        \tgen{C} & $0.65$ & $0.58$ & $0.65$ & \\
        \cline{2-4}
        \tgen{letters} & \tgen{A} & \tgen{B} & \tgen{A} & \tgen{$\dots$} \\
      \end{tabular}}\\[2pt]
      $\rightarrow$ k-means: clusters \tgen{$\upalpha$}, \tgen{$\upbeta$}, ...\\[3pt]
      then k-means on the numbers};

    \node[font=\scriptsize\itshape, fill=white, inner sep=1pt] (orlab)
      at ($(sage.east)!0.5!(sagetwo.west)$) {or};

    \node[wfout, anchor=north] (cl1) at (12.15,-3.8) {%
      \begin{tabular}{@{}c@{\ \,}l@{}}
        \wfplot{1.8pt}{(0,0)(1,6)(2,0)(3,6)(4,0)(5,6)(6,0)(7,6)(8,0)} & cluster \tgen{$\upalpha$} \\[2pt]
        \wfplot{1.8pt}{(0,0)(1,1)(2,3)(3,6)(4,8)(5,6)(6,3)(7,1)(8,0)} & cluster \tgen{$\upbeta$} \\[1pt]
        \multicolumn{2}{@{}c@{}}{$\dots$} \\
      \end{tabular}};
    \node[wfout, anchor=north] (cl2) at (15.95,-3.8) {%
      \begin{tabular}{@{}c@{\ \,}l@{\quad}c@{\ \,}l@{}}
        \wfplot{1.8pt}{(0,0)(1,3)(2,0)(3,3)(4,0)(5,3)(6,0)(7,3)(8,0)} & \tgen{$\upalpha$1} &
        \wfplot{1.8pt}{(0,0)(1,6)(2,0)(3,6)(4,0)(5,6)(6,0)(7,6)(8,0)} & \tgen{$\upalpha$2} \\[2pt]
        \wfplot{1.8pt}{(0,0)(1,0.5)(2,1.5)(3,3)(4,4)(5,3)(6,1.5)(7,0.5)(8,0)} & \tgen{$\upbeta$1} &
        \wfplot{1.8pt}{(0,0)(1,1)(2,3)(3,6)(4,8)(5,6)(6,3)(7,1)(8,0)} & \tgen{$\upbeta$2} \\[1pt]
        \multicolumn{4}{@{}c@{}}{$\dots$} \\
      \end{tabular}};

    \draw[wfarr] (genomes.south) -- (resample.north);
    \draw[wfarr] (resample.south) -- ++(0,-0.25) -| (sage.north);
    \draw[wfarr] (resample.south) -- ++(0,-0.25) -| (sagetwo.north);
    \draw[wfarr] (sage) -- (cl1);
    \draw[wfarr] (sagetwo) -- (cl2);

    \coordinate (tA) at (2.9,1.05);   \coordinate (bA) at (tA |- cl2.south);
    \coordinate (tB) at (8.4,1.05);   \coordinate (bB) at (tB |- cl2.south);
    \coordinate (tC) at (13.95,1.05); \coordinate (bC) at (tC |- cl2.south);
    \begin{scope}[on background layer]
      \node[wfpanel, fit=(tA)(bA)(traces)(shapekm)(letterA)(letterB)(ampA)(ampB)(symA)(symB)(znote)] (p1) {};
      \node[wfpanel, fit=(tB)(bB)(onetrace)(w1)(w2)(w3)(genome)] (p2) {};
      \node[wfpanel, fit=(tC)(bC)(genomes)(resample)(sage)(sagetwo)(orlab)(cl1)(cl2)] (p3) {};
    \end{scope}
    \node[wftitle, inner sep=1pt, text depth=0.25em, above=0pt of p1.north] {1. Alphabet learning};
    \node[wftitle, inner sep=1pt, text depth=0.25em, above=0pt of p2.north] {2. Genome translation};
    \node[wftitle, inner sep=1pt, text depth=0.25em, above=0pt of p3.north] {3. Clustering};
  \end{tikzpicture}
  \caption{\sage workflow.
  The \textbf{alphabet learning} pools windows of length $W$ from all traces and infer an alphabet based on their shape (letters) and amplitude (numbers).
  The \textbf{genome translation} translates each trace into a sequence of symbols from the alphabet.
  The \textbf{clustering} normalizes every genome to $L$ symbols and replaces each symbol by how close it is to the others, then simply clusters the traces (\sage), or clusters them first on the letters and then on the numbers (bandwidth levels) inside each shape cluster (\sagetwo).
  \sagetwo clusters are named with a Greek letter and a level (for instance \tgen{$\upalpha$1}), so a cluster is never confused with an alphabet symbol (like \tgen{A1}).}
  \label{fig:workflow}
\end{figure*}

\sage therefore consists of three steps (see \fig~\ref{fig:workflow}): learning the alphabet, translating each trace using this alphabet, and clustering the resulting representations.

\subsection{Alphabet Learning}

\begin{figure}[t]
  \centering
  \includegraphics[width=\linewidth]{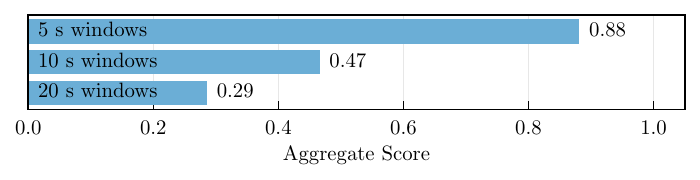}
  \caption{Aggregate score (see Section~\ref{sec:metrics}) for \sagetwo with three $W$ values. Write traces of \plafrim. Higher is better.}
  \label{fig:window-comparisons}
\end{figure}

\sage first learns an alphabet from the set of traces. The goal of this stage is to identify recurring I/O patterns.
We therefore pool all traces from the data set and divide them into non-overlapping windows of $W$ consecutive samples.
We use $W=5$, which gives the best results on \plafrim (\fig~\ref{fig:window-comparisons}).
Clustering techniques combine shape and amplitude, allowing one to potentially dominate the other. To avoid this, we distinguish for each window: its \emph{shape}, which captures the temporal variation of its I/O activity, and its \emph{amplitude}, which captures its absolute bandwidth level.
The alphabet is therefore learned in two stages, as illustrated in \fig~\ref{fig:workflow} (left).

\paragraph{Shape clustering}
The goal is to ensure that windows with the same up-and-down pattern are assigned to the same shape cluster. To do so, we first z-normalize each window by subtracting its mean and dividing by its standard deviation, removing differences in absolute level and scale. We then compute the differences between consecutive points to get the shape vector of the window.
\begin{rronly}
Let $x = (x_1, \dots, x_W)$ be a window with mean $\mu_x$ and standard deviation $\sigma_x$.
We z-normalize the window and take the differences between consecutive points, which gives the shape vector of the window:
\begin{equation}
  sv_t = \frac{x_{t+1} - \mu_x}{\sigma_x} - \frac{x_t - \mu_x}{\sigma_x}, \qquad t = 1, \dots, W-1,
\end{equation}
\end{rronly}
Z-normalization removes the absolute scale of a window, so the same pattern at $1$~MiB/s and $1$~GiB/s produces the same shape vector. Consequently, two windows are considered similar when their I/O activity rises and falls at the same time steps, rather than simply reaching similar values.
We then cluster (\kmeans) the shape vectors.
We select the number of clusters between $2$ and $50$ using an elbow search based on the Kneedle~\cite{kneedle} algorithm.
\begin{rronly}
The elbow search first evaluates every fifth value of $k$, then every value around the knee found in this coarse pass.
It runs on a random sample of at most two million windows, with 3 \kmeans initializations per value of $k$.
The final models use 20 initializations.
\end{rronly}

\begin{figure}
  \centering
  \includegraphics[width=0.32\linewidth]{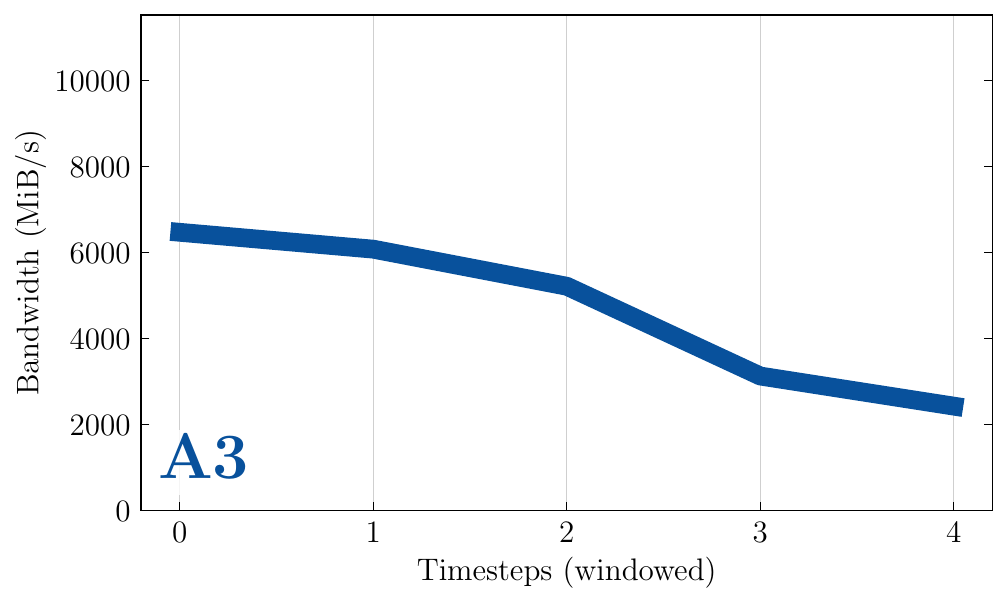}\hfill
  \includegraphics[width=0.32\linewidth]{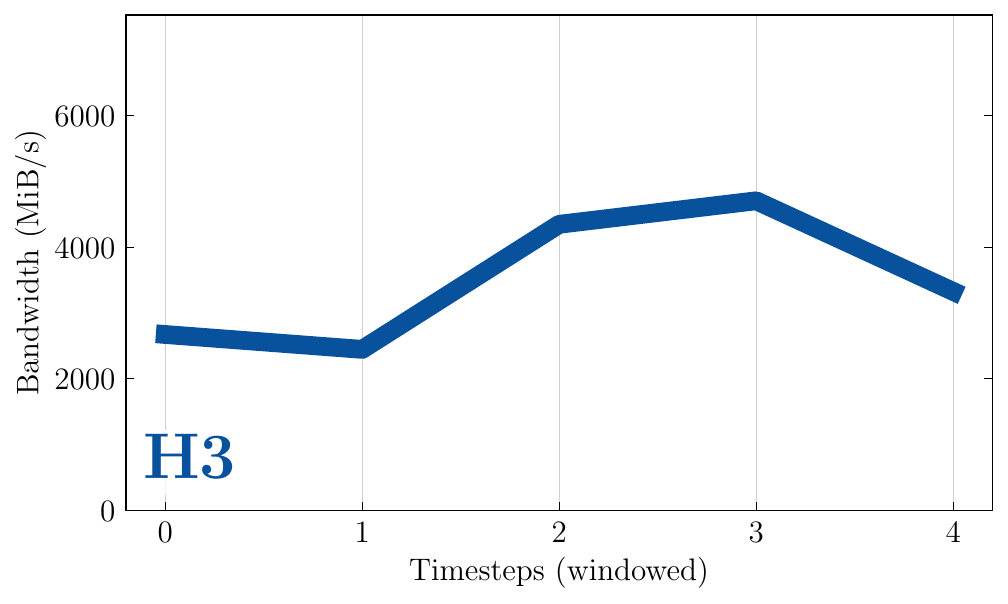}\hfill
  \includegraphics[width=0.32\linewidth]{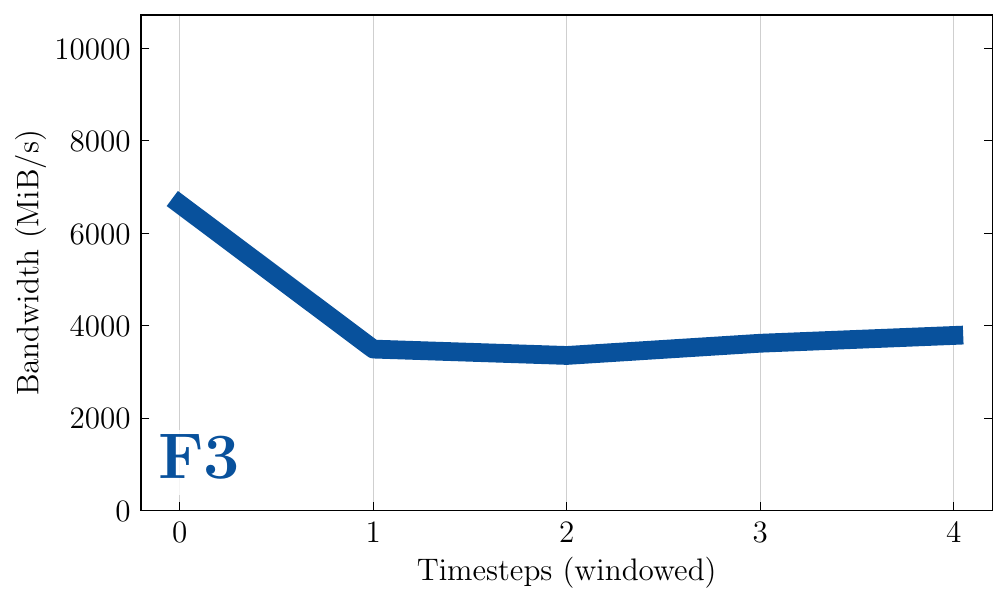}
  \includegraphics[width=0.32\linewidth]{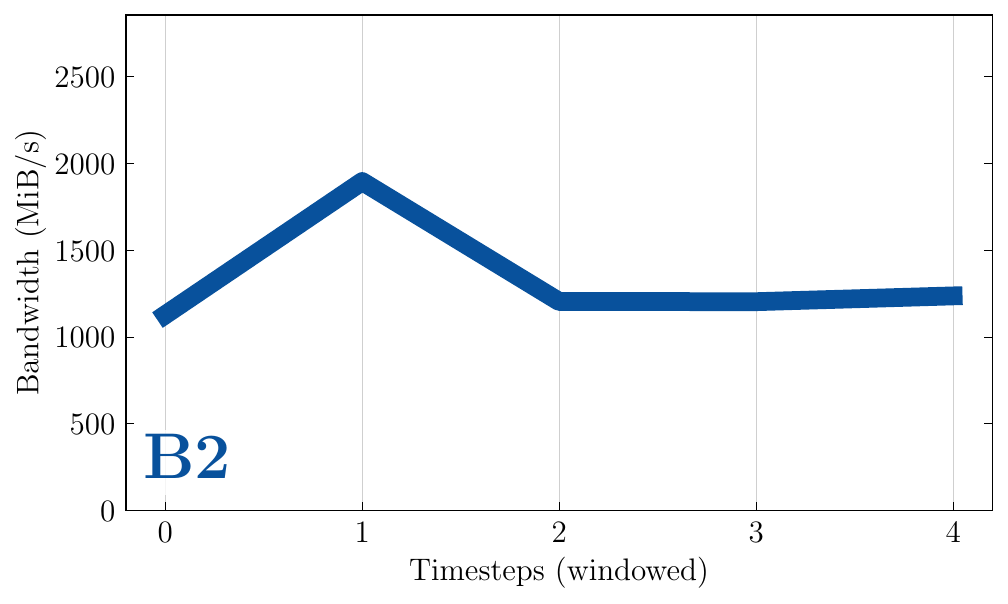}\hfill
  \includegraphics[width=0.32\linewidth]{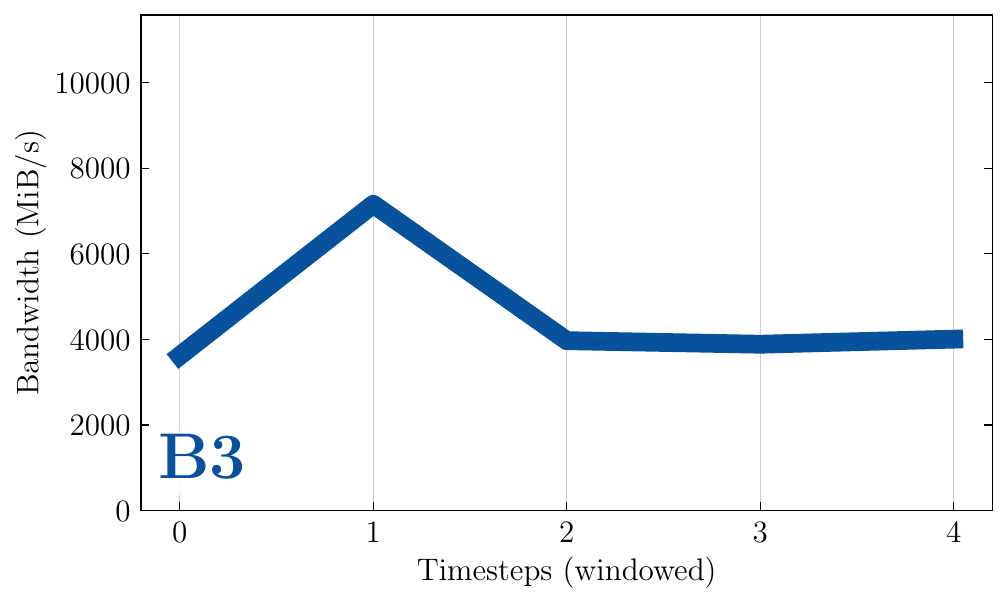}\hfill
  \includegraphics[width=0.32\linewidth]{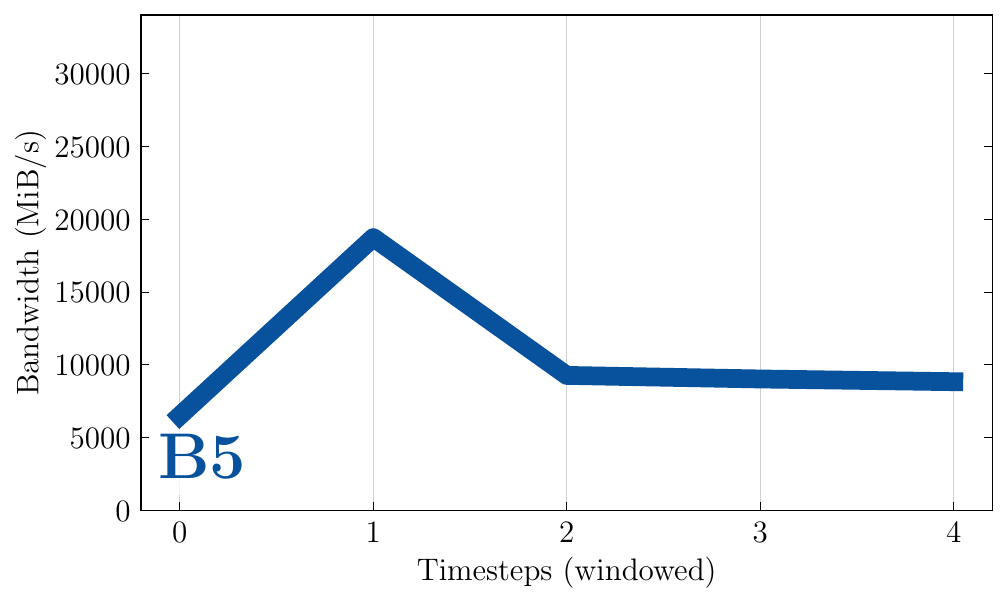}
  \caption{Example of an alphabet built by \sage.
  Top row: three shapes (\tgen{A}, \tgen{H}, \tgen{F}), each at its third amplitude level.
  Bottom row: shape \tgen{B} at levels \tgen{2}, \tgen{3} and \tgen{5}.
  The y-axes are all different.}
  \label{fig:genome}
\end{figure}

\paragraph{Amplitude clustering}
Within each shape cluster, we apply a second clustering step to distinguish windows with different amplitude levels.
The number of amplitude sub-clusters is selected using an elbow search over all windows and is shared across all shape clusters.
The amplitude sub-clusters are labeled with numbers, and a higher number always means a higher bandwidth within the letter, as illustrated by \tgen{B2}, \tgen{B3}, and \tgen{B5} in \fig~\ref{fig:genome}.
The amplitude centroids are learned independently for each shape. Thus, \tgen{A3} and \tgen{H3}, for example, both denote the third amplitude level of their respective shapes, but they do not necessarily correspond to the same absolute bandwidth.
A window without any I/O activity is assigned the symbol \tgen{Z1}.
The resulting set of letter-number pairs (that we call symbols) constitutes the learned alphabet.

\subsection{Genome Translation}
To represent each trace as a sequence of recurring I/O patterns, we translate it window by window, first identifying the nearest shape centroid and then the nearest amplitude centroid within that shape, as illustrated in \fig~\ref{fig:workflow} (middle).
The time-ordered sequence of symbols obtained from a trace is what we call its ``genome''.
The ``genome'' provides a lightweight representation of the original trace while preserving the recurrence and temporal ordering of its local patterns.

\subsection{Clustering}\label{sec:sage_clustering}

Finally, we create clusters from the genomes (right of \fig~\ref{fig:workflow}).
The intuition is that two jobs with similar I/O profiles should have genomes containing similar symbols at similar relative positions. We therefore compare genomes position by position. For this comparison to be meaningful, genomes must have the same length, motivating the normalization step. In addition, directly comparing symbols would treat all distinct symbols as equally different. We therefore represent each symbol by a vector of its similarities to the other symbols.

Because our symbols encode both shape and amplitude, we propose two variants.
\sage clusters genomes in a single step using the full symbols (so letters and numbers).
This representation primarily groups jobs according to when and with which shape they perform I/O, while amplitude has a smaller influence.
\sagetwo instead first groups jobs according to the shapes in their genomes and then splits each shape group by amplitude level.
\sagetwo clusters are therefore able to capture both when a job performs I/O and at which bandwidth, but at the cost of producing more clusters, some of which may be very similar.

\paragraph{Normalization}
The genome length is normalized to $L$ symbols.
A genome of $n \leq L$ symbols is stretched to $L$ symbols by nearest-neighbor interpolation, so each of its symbols is repeated about $L/n$ times.
The ordering of the symbols is preserved, so symbol $s_i$ at the $i$-th position will be at the $\frac{i \times L}{n}$-th position of the genome.
That way, we capture the proportion of a job's duration that had elapsed before a given symbol.
For instance, the genome \tgen{A2 B3 B3 A3} brought to $L = 8$ positions becomes \tgen{A2 A2 B3 B3 B3 B3 A3 A3}.
A longer genome is split into $L$ bins of equal length.
For each bin, the centroids of its symbols are concatenated and resampled to $W$ points by interpolation.
The bin is then given the closest symbol from the alphabet. 

\paragraph{Similarity vectors}
With the default clustering distances, every pair of distinct symbols would be at the same distance, so \tgen{A1} and \tgen{A2} would be as distant as \tgen{A1} and \tgen{K3} for example.
Instead, we replace each symbol $s_i$ by a vector that encodes how close it is to every other symbol, and these vectors are then used for clustering as explained below.

\paragraph{\sage clustering}

The similarity of two symbols, $Sim_{s_i,s_j}$, is the cosine similarity of their centroids. 
\tgen{Z1} is similar to no other symbol.
Each symbol of each genome is replaced by the vector of its similarities to all other symbols of the alphabet.
For instance, as shown in \fig~\ref{fig:workflow}, \tgen{A2} becomes $(Sim_{\mathtt{A2},\mathtt{A1}} = 0.97, Sim_{\mathtt{A2},\mathtt{A2}} = 1, ...)$.
A final clustering step (\kmeans with the elbow method) gives the clusters for \sage.

\paragraph{\sagetwo clustering}

\sagetwo runs the clustering step twice, first on shape and then on amplitude, like alphabet learning does for windows.
In the first stage, we cluster the genomes, but we only keep the letters of the symbols, so \tgen{A1}, \tgen{A2}, \tgen{A3}, $\dots$ all become \tgen{A}.
Each letter is replaced by the vector of its similarities to other letters, computed between the shape centroids, so two letters with a similar up-and-down pattern are close in similarity.
In the second stage, we cluster the genomes \emph{from each shape cluster} by keeping only the numbers, and each number $n$ is replaced by a vector of $n-1$ ones followed by zeros.
We choose to do this because, since \kmeans uses the squared Euclidean distance, using the numbers directly would square their difference.
To avoid this, we build vectors that represent linear distances. 
For example, with three amplitude levels, \tgen{1} becomes $(0,0,0)$, \tgen{2} becomes $(1,0,0)$ and \tgen{3} becomes $(1,1,0)$, so \tgen{1} and \tgen{3} are at squared distance $2$.
The final clusters are named with a Greek letter for the shape and a number for the level, for instance \tgen{$\upalpha$2} for the second amplitude cluster of shape cluster \tgen{$\upalpha$}.

If a cluster has fewer than five traces, it is merged with the cluster that has the closest centroid (for \sagetwo, the closest within the same shape class).
Note that our approach uses \kmeans for clustering, with the Euclidean distance for comparisons and the elbow method to select the number of clusters. Nonetheless, \sage is not specific to these techniques, and other clustering techniques could be applied to the workflow depicted in \fig~\ref{fig:workflow}.

\subsection{\sage in Practice}

\fig~\ref{fig:sagetwo-classes-write} shows the level-\tgen{3} clusters that \sagetwo builds from the \plafrim write traces.
\tgen{$\upbeta$3} writes throughout the run, \tgen{$\upepsilon$3} at the end, \tgen{$\updelta$3} at the beginning and in the middle, and \tgen{$\upzeta$3} writes more and more as the job progresses.
A scheduler can, for example, use this to allocate jobs that write at the same moment to different I/O resources.
Time series of I/O activity (read/write/metadata) over time can be collected by a profiling tool such as Darshan~\cite{original_darshan_paper, darshan}, Recorder~\cite{recorder}, or an online collection tool such as TOTO~\cite{boito:hal-05662894}. \sage can be used periodically to identify/update the classes. A job will belong to multiple classes, one for each metric being monitored (write/read/etc.). Then, at execution time, a trace from a previous execution of the same application could be used to find its class~\cite{9460487,7877148}, or one could predict its class from job submission information~\cite{9820667} and/or from a partial time series~\cite{betke2021classifying}. Then, an I/O scheduler can use the classes' and jobs' information to make decisions (either online or offline). We plan on studying the prediction of classes as future work, since our focus here is clustering I/O time series. 

\begin{figure}[t]
  \centering
  \includegraphics[width=\linewidth]{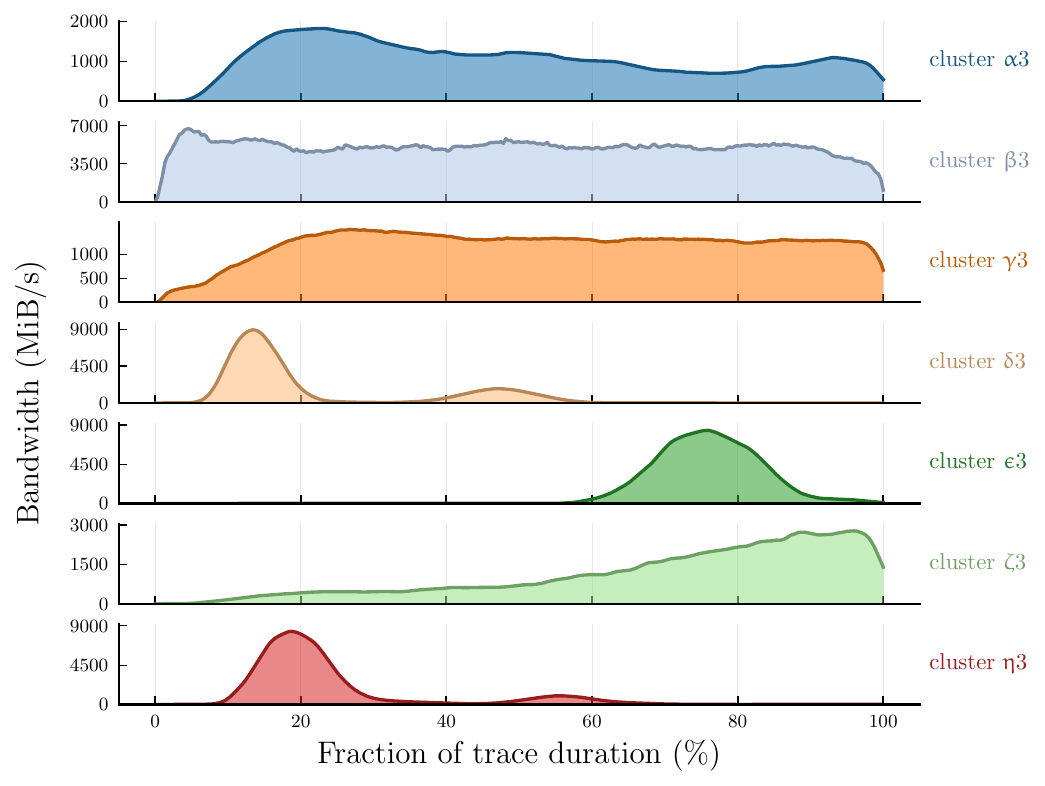}
  \caption{Centroids of the level-\tgen{3} clusters built by \sagetwo on the \plafrim write traces.
  Only level \tgen{3} of each shape cluster is shown, for readability.}
  \label{fig:sagetwo-classes-write}
\end{figure}

\section{Baselines and Workload}\label{sec:baseline_and_workload}

This section describes the clustering techniques and the data sets we use.
Both serve the clustering quality evaluation of Section~\ref{sec:evaluation} and the scheduling use case of Section~\ref{sec:simulation}.

\subsection{Clustering Baselines}\label{sec:baselines}

We implement \sage in Python with scikit-learn,
and compare it to other commonly used techniques. 
All of them except \tsfresh first resample each trace to $L$ points by linear interpolation ($L=300$ by default; Section~\ref{sec:results_length} shows that $L$ has little impact).
As for \sage (Section~\ref{sec:sage_clustering}), we merge every cluster of fewer than five traces into the cluster with the closest centroid, so that a single outlier does not form a cluster of its own.

\paragraph{\kmeans}
combined with the Euclidean distance to compare the points, is widely used for clustering I/O information in HPC~\cite{10.1145/3391812.3396270, pavan2019unsupervised,bang2021workflow,betke2021classifying,Halawa_2020}.
\begin{rronly}
It is a centroid-based algorithm that aims to minimize the distance (in this case squared Euclidean) between the points and their associated centroid. A big issue often raised is that the number of clusters is a parameter that needs to be set by the user. We use an elbow method to determine the optimal number of clusters while sweeping between 2 and 50.
The elbow curve is the inertia of the fitted model, that is the sum over all points of the squared Euclidean distance to their centroid, which decreases as $k$ grows. We locate its knee with the Kneedle algorithm~\cite{kneedle, kneed}. 
\end{rronly}
\begin{cponly}
We used the elbow method to select the number of clusters between 2 and 50, using the Kneedle algorithm~\cite{kneedle,kneed}.
\end{cponly}

\paragraph{\kshape} is another clustering algorithm~\cite{kshape_funding_paper,new_shabe_based_metric} that aims at optimizing a distance metric called SBD~\cite{sbd_funding_paper,sbd}, which tries to give more importance to the shape of curves. 
\begin{rronly}
\kshape z-normalizes the bandwidth values of each trace as its distance computation requires it.
\kshape proposes a cluster center determination strategy based on matrix decomposition.
\end{rronly}
We used the tslearn implementation~\cite{kshape}.

\paragraph{\tsfresh} from the Python package of the same name, extracts features from time series~\cite{tsfresh_founding_paper, tsfresh,tsfresh_used_for_hydro_clustering,tsfresh_used_for_geothermal_labelling}.
\begin{rronly}
It is widely used for systematic feature engineering from time series and other sequential data.
For instance, other works use \tsfresh to extract features from time series and then use clustering to group the extracted features in the domains of Water level data \cite{tsfresh_used_for_hydro_clustering} or Geothermal data \cite{tsfresh_used_for_geothermal_labelling}.
\end{rronly}
It extracts a large number of features and then uses a feature selection method to select the most relevant ones for the clustering task (we used the EfficientFCParameters method~\cite{EfficientFCParameters}). We then use \kmeans for clustering on the extracted features.

\paragraph{\kmeans + Dynamic Time Warping (DTW)} uses DTW, from tslearn~\cite{dtw}, to compute the distance between the time series~\cite{DTW_used_for_clustering,A_Deep_Look_into_the_Temporal_IO_Behavior}. DTW is considered a suitable choice for shape-based clustering as it matches the series by stretching or compressing segments, allowing it to compare patterns that occur at different time periods.
\begin{rronly}
In our case we did not constrain the left right lookup.
\end{rronly}


\begin{rronly}
\paragraph{User-based classification} This baseline groups jobs by user ID, which is the only additional information available for the \plafrim data set. In this comparison, it represents approaches that assume uniform behavior across jobs from the same user~\cite{9820667}.
\end{rronly}

\subsection{Data Curation and Pre-processing} \label{sec:methodology_data}

Techniques are compared using \numberofdatasets data sets, which represent all jobs executed on a system in a certain time period. 

\paragraph{\plafrim} 
is an experimental platform 
composed of 192 nodes and a BeeGFS file system. 
\ifdoubleblind
This data set, composed of $108{,}433$ jobs, covers a period from May 2022 to June 2026.  
\fi 
Data was collected with the \texttt{beegfs-ctl} command, called every second.
We make this data set publicly available.\footnote{It will be available in the camera-ready version of the paper.}

\paragraph{\polaris}
\ifdoubleblind
is a 520-node production supercomputer with a Lustre file system.
This publicly available data set, composed of $860{,}006$ Darshan traces, covers all of 2025.
\fi
Darshan's heatmap module records bytes read and written in up to 200 time bins, distributing I/O operations across bins proportionally to their duration.

\paragraph{Filtering}
First, we discard jobs that contain too few samples for meaningful patterns: jobs shorter than 30 seconds on \plafrim (47.4\%) and jobs with fewer than 20 bins on \polaris (9.1\%).
On \plafrim, we discard jobs longer than 3 hours (2.6\%), which are rare and expensive to process for slow techniques such as \kmeansdtw.
Second, we set all samples below 1~MiB/s to 0 to decrease noise.
\begin{rronly}
This sets $86.6\%$ of the $73.8$~million remaining samples to zero on \plafrim, and $97.1\%$ of the $130$~million remaining samples on \polaris.
\end{rronly}
Third, we discard time series with fewer than 5 non-zero samples on \plafrim, and time series without any non-zero sample on \polaris.
Read and write traces are filtered independently, so, for example, the read time series of a job could be retained while its write time series is discarded.
In the end, we obtain $28{,}281$ read and $40{,}208$ write traces on \plafrim, and $582{,}316$ read and $168{,}966$ write traces on \polaris.

\section{Clustering Quality Evaluation}\label{sec:evaluation}

This section evaluates the quality of the clustering, using the baselines and workloads presented in Section~\ref{sec:baseline_and_workload} and the metrics presented in Section~\ref{sec:metrics}.
Section~\ref{sec:results_score} presents the aggregate score results, Section~\ref{sec:results_time} the clustering time, and Section~\ref{sec:results_length} the impact of trace resolution.

\subsection{Metrics}\label{sec:metrics}

A good clustering groups jobs that are similar, and separates jobs that are not. Hence, we compute each metric twice: i) the \intra value is the distance from a trace to the centroid of its cluster, and ii) the \inter value is the distance between the centroids of every pair of clusters.
The goal is to achieve low \intra and high \inter values.
We report the median values.

Note that we are interested in characterizing jobs not only by their amplitude (i.e., their I/O bandwidth), but also by their shape (i.e., when they do I/O). In this context, simple metrics such as the Euclidean distance are dominated by the amplitude and fail to capture differences in shape.
So, we present here two shape-oriented metrics from the literature, \ddtw and \sbd.
We then propose a new metric, \pld, which addresses shortcomings of \ddtw and \sbd by accounting for both when a job performs I/O and at which bandwidth.

\paragraph{Derivative Dynamic Time Warping (DDTW)~\cite{DDTW}}
\begin{cponly}
is a variant of DTW which compares the derivatives of the time series, focusing on differences in their slopes and temporal variations.
\end{cponly}
\begin{rronly}
is a variant of DTW, widely used for measuring the distance between time series while accounting for temporal shifts. However, DTW can compensate for differences in amplitude by warping the time axis, potentially considering time series with similar shapes as very different.
DDTW instead compares the derivatives of the time series, thereby focusing on differences in their slopes and temporal variations. This better matches the notion of I/O shape difference that we aim to capture.
\end{rronly}
To reduce the cost of DDTW, we restrict the warping window to sixty points on either side of each point.

\paragraph{Shape-Based Distance (SBD)}
is a distance metric~\cite{sbd} that 
 focuses on the shape of time series rather than their absolute amplitude~\cite{sbd_funding_paper, new_shabe_based_metric}.

\begin{figure}[t]
  \centering
  \includegraphics[width=1\linewidth]{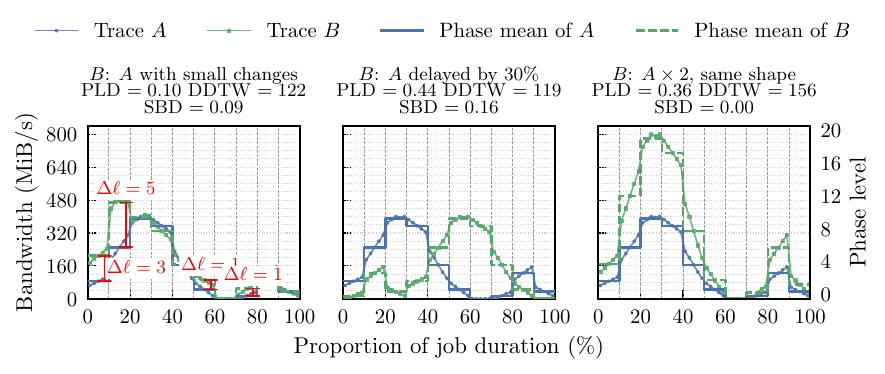}  
  \caption{Example of \pld usage\rrtext{ with the traces of Table~\ref{tab:phase-level}}.
  Markers are the samples, and the thick steps are the phase means, from which the levels of \pld are computed.
  Dashed vertical lines separate the $P$ phases and dotted horizontal lines separate the $\ell$ levels.
  Red capped lines (left fig) mark the phases that land in different levels\rrtext{, i.e.\ the last row of Table~\ref{tab:phase-level}}. \pld is the sum of these differences divided by $P \times \ell$.
  }
  \label{fig:phase-level}
\end{figure}

\paragraph{Phase-Level Distance (\pld)}
\sbd ignores amplitude, and \ddtw warping can lead to uneven results.
\fig~\ref{fig:phase-level} shows this on a trace $A$ and three versions of a trace $B$: $A$ with small changes, $A$ delayed by $30\%$ of the job duration, and $A$ scaled twofold.
Intuitively, the second and third cases should yield worse metric values than the first.
\sbd ignores the amplitude of both series, so it scores the scaled $B$ at exactly $0$, as if it were $A$ itself.
\ddtw warps the time axis to absorb the delay, and ranks the delayed $B$ ($119$) closer to $A$ than the slightly changed one ($122$).
We therefore propose \pld, the only one of the three metrics that ranks the slightly changed $B$ closest ($0.10$), ahead of the scaled ($0.36$) and the delayed ($0.44$) ones.

\pld cuts the time axis into phases and the bandwidth axis into levels, and checks whether $B$ sits in the same level as $A$ in every phase.
Both series, resampled to length $L$, are cut into $P$ phases of $L/P$ points.
Let $\bar{a}_p$ and $\bar{b}_p$ be the mean bandwidth of $A$ and $B$ in phase $p$, and $A_{\max}$ the peak bandwidth of $A$.
A level is $A_{\max}/\ell$ wide, so the levels follow the scale of $A$, whether it peaks at thousands or at tens of MiB/s:
\begin{equation}\label{eq:phase-level}
  \pld(B,A)=\frac{1}{P \times \ell}\sum_{p=1}^{P}\left|\left\lfloor\frac{\ell\,\bar{b}_p}{A_{\max}}\right\rfloor-\left\lfloor\frac{\ell\,\bar{a}_p}{A_{\max}}\right\rfloor\right|
\end{equation}
Averaging within a phase absorbs small time shifts, and the levels absorb small bandwidth differences.
Levels are not capped, so a $B$ above the peak of $A$ is still penalized in proportion.
A \pld of $0$ means $B$ sits in the level of $A$ in every phase, and $0.1$ that it is one level away on average.
\begin{rronly}
Table~\ref{tab:phase-level} details the computation for the left panel of \fig~\ref{fig:phase-level}.
\end{rronly}
We use $P=10$ phases and $\ell=10$ levels by default in the paper.

\begin{rronly}
\begin{table*}[t]
  \centering
  \caption{\pld between a synthetic trace $A$ and a synthetic trace $B$, with $P = 10$ phases and $\ell = 10$ levels.
  Each trace has $50$ samples, five per phase, and the table reports their phase means.
  $A$ peaks at $A_{\max} = 400$~MiB/s, which means that one level is $40$~MiB/s.
  Note that on phase~2 $B$ exceeds the peak of $A$ and lands in level $11$.
  Summing the last row gives $\pld(B,A) = 10 / (P\ell) = 0.10$: over the whole duration, $B$ sits on average $10\%$ of the peak bandwidth of $A$ away from $A$.}
  \label{tab:phase-level}
  \begin{tabular}{lcccccccccc}
    \toprule
    Phase $p$ & 1 & 2 & 3 & 4 & 5 & 6 & 7 & 8 & 9 & 10 \\
    \midrule
    Phase mean $\bar{a}_p$ (MiB/s) & 85 & 250 & 390 & 355 & 165 & 45 & 0 & 15 & 125 & 35 \\
    Level of $A$                 & 2 & 6 & 9 & 8 & 4 & 1 & 0 & 0 & 3 & 0 \\
    \addlinespace
    Phase mean $\bar{b}_p$ (MiB/s) & 210 & 468.2 & 395 & 330 & 170 & 90 & 0 & 50 & 130 & 35 \\
    Level of $B$                 & 5 & 11 & 9 & 8 & 4 & 2 & 0 & 1 & 3 & 0 \\
    \midrule
    Level gap                    & 3 & 5 & 0 & 0 & 0 & 1 & 0 & 1 & 0 & 0 \\
    \bottomrule
  \end{tabular}
\end{table*}
\end{rronly}

\paragraph{Cluster balance} We observed that some clustering techniques have a tendency to make clusters of a single job or very few jobs, which skews the metrics. Hence, we add the normalized Shannon entropy~\cite{shannon1948} to represent the balance in cluster size.
\begin{rronly}
To do so, we convert the number of traces per cluster into probabilities by dividing each count by the total number of traces.
Then, we compute the Shannon entropy, and finally, we normalize its value by the maximum possible entropy.
\end{rronly}
The closer the value is to 1, the more evenly the traces are split across the clusters.

\begin{rronly}
In more detail:
Let $K$ be the number of non-empty clusters, $n_i$ the number of traces in cluster $i$, and $N = \sum_{i=1}^{K} n_i$ the total number of traces.
We compute the cluster balance as:
\begin{equation}
  \label{eq:balance}
  B = -\frac{1}{\log_2 K} \sum_{i=1}^{K} p_i \log_2 p_i .
\end{equation}
Dividing by $\log_2 K$ lets us normalize the entropy.
\end{rronly}

\paragraph{Aggregate score}
To compare strategies using a single value, we combine the median \intra and \inter values of the previous distance metrics and the cluster balance metric into an aggregate score.
Each metric is then min-max normalized across the compared strategies to $[0, 1]$, with $1$ given to the best strategy and $0$ to the worst.
The aggregate score of a strategy is the mean of the normalized metrics.

\subsection{Aggregate Score Results}\label{sec:results_score}

\begin{figure}[t]
  \centering
  \begin{subfigure}[t]{\linewidth}
    \centering
    \includegraphics[width=\linewidth]{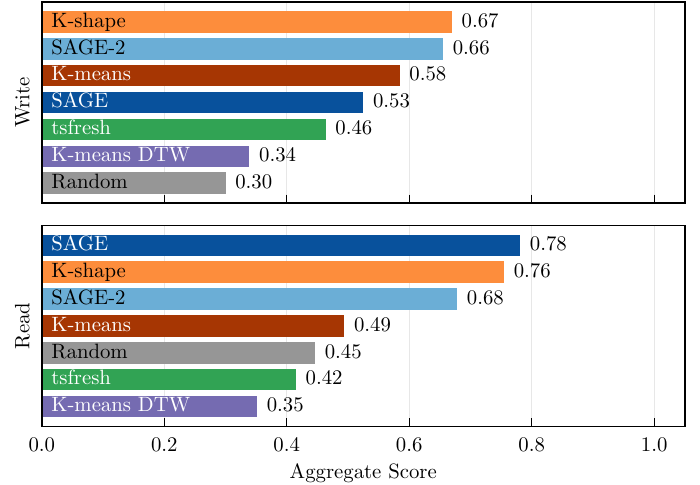}
    \caption{\plafrim.}
    \label{fig:aggregate}
  \end{subfigure}
  \\[0.5em]
  \begin{subfigure}[t]{\linewidth}
    \centering
    \includegraphics[width=\linewidth]{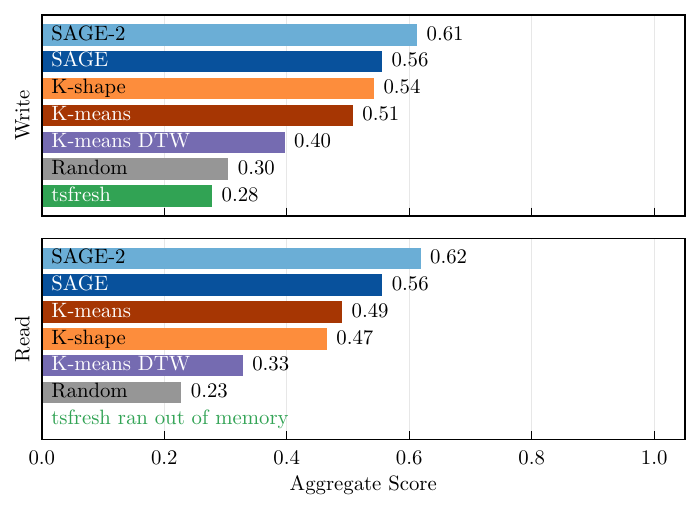}
    \caption{\polaris.}
    \label{fig:aggregate-polaris}
  \end{subfigure}
  \caption{Aggregate score of all metrics for each data set, write (top) and read (bottom). \rrtext{Min-max normalized across strategies.} Higher is better.}
  \label{fig:aggregate-all}
\end{figure}

\begin{figure}[t]
  \centering
  \includegraphics[width=1\linewidth]{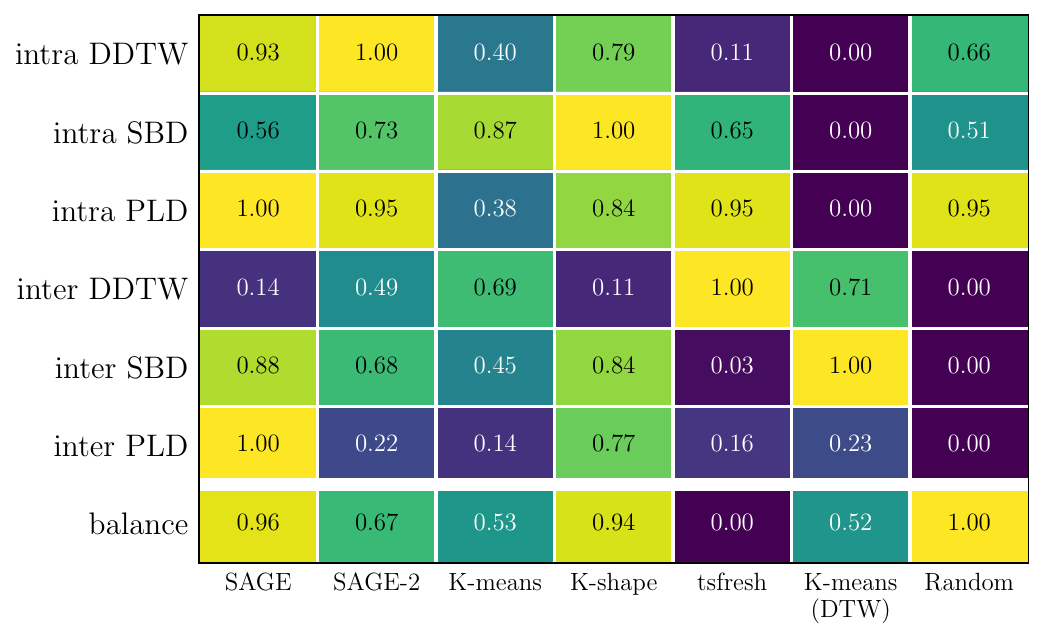}
  \caption{Breakdown of the \plafrim read scores of \fig~\ref{fig:aggregate} per metric. Each cell is the median of the metric over every trace. $1$ is the best strategy for that metric and $0$ the worst.}
  \label{fig:aggregate-heatmap-read}
\end{figure}

\begin{figure}[t]
  \centering
  \includegraphics[width=\linewidth]{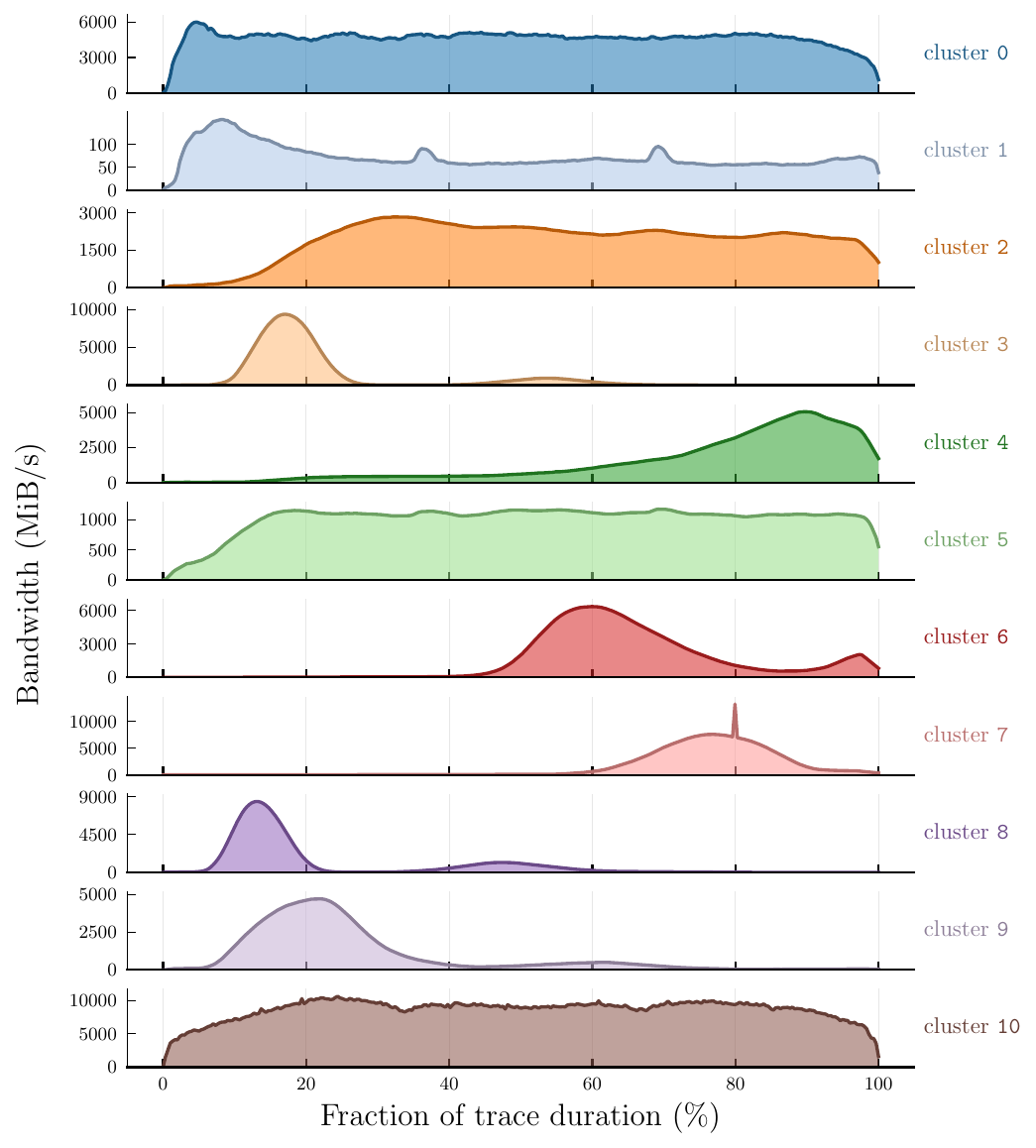}
  \caption{Clusters produced by \kmeans on \plafrim. Write.}
  \label{fig:kmeans-classes-write}
\end{figure}

\fig~\ref{fig:aggregate-all} summarizes the clustering results on \plafrim (\fig~\ref{fig:aggregate}) and \polaris (\fig~\ref{fig:aggregate-polaris}).
\sage or \sagetwo obtain the highest aggregate score in three of the four cases.
On \plafrim' writes, \kshape is within $0.01$ of \sagetwo, but takes on average $54$ times longer across both systems
(Section~\ref{sec:results_time}).
\kmeansdtw, \tsfresh, and \random take the last three places on both systems, and \tsfresh runs out of memory on \polaris read traces.
\sagetwo does well because its hierarchical construction produces many homogeneous clusters that cover a broad range of I/O behaviors and amplitude levels\rrtext{ (\fig~\ref{fig:sagetwo-classes-write})}, so each trace finds a cluster that closely matches its activity.
\kmeans, in contrast, produces $11$ clusters on \plafrim, and two of them, both constant-bandwidth plateaus, gather $64\%$ of the traces (clusters 1 and 5 in \fig~\ref{fig:kmeans-classes-write}).
Behaviors that \sagetwo separates are thus merged by \kmeans, which increases their \intra distance.

\begin{figure}[t]
  \centering
  \includegraphics[width=\linewidth]{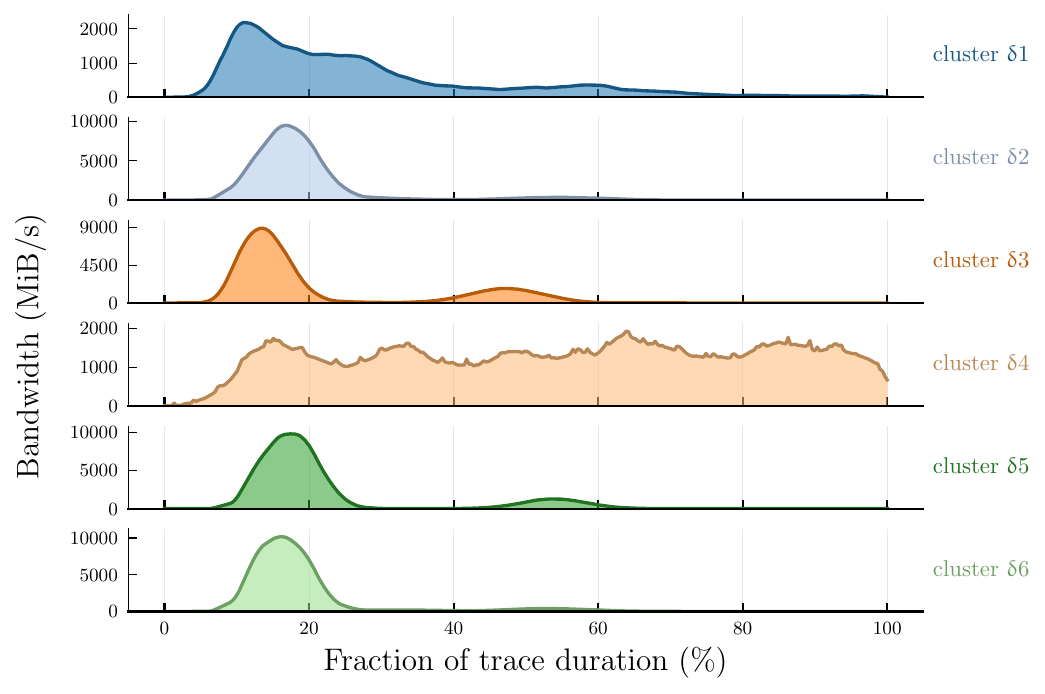}
  \caption{\tgen{$\updelta$} clusters produced by \sagetwo on \plafrim. Write.}
  \label{fig:sagetwo-classes-write-delta}
\end{figure}

The heatmap in \fig~\ref{fig:aggregate-heatmap-read} breaks down the \plafrim read scores per metric.
\sagetwo obtains the best \intra \ddtw and \sage the best \intra \pld, meaning that their cluster centroids closely represent their members.
Only \kshape, which optimizes \sbd directly, also does well on \intra metrics.
Looking at the \inter scores, \sage performs well on both \sbd and \pld, while \sagetwo obtains intermediate scores.
This is expected from its hierarchical construction, which creates multiple classes with similar shapes but different amplitude levels, reducing separation between clusters.
\fig~\ref{fig:sagetwo-classes-write-delta} illustrates this effect for the write activity of the \tgen{$\updelta$} classes, whose centroids share similar shapes.
It also shows a limitation of \sagetwo clustering.
We observe that \tgen{$\updelta$4} does not retain the same shape as the other \tgen{$\updelta$} classes.
This does not mean that its members have a different shape.
Rather, averaging their individual windows can smooth or distort the shape identified during the first clustering stage.
Similarly, \tgen{$\updelta$3}, \tgen{$\updelta$5} and \tgen{$\updelta$6} have only small differences in peak amplitude.
Such small differences can nevertheless be sufficient to separate clusters, for example when the height of a secondary peak or the proportion of zero-I/O windows differs.
These observations suggest that the hierarchical construction could be refined by manually tuning the number of amplitude levels per shape. We deliberately avoid such specialization so that \sagetwo remains fully automated and independent of the workload.

\tsfresh achieves the best \inter \ddtw score but the worst cluster balance, with nearly all traces assigned to a single cluster. The few traces assigned to the other clusters are therefore very different from the dominant cluster, resulting in large \inter \ddtw distances.

\begin{rronly}
\begin{figure}[t]
  \centering
  \includegraphics[width=\linewidth]{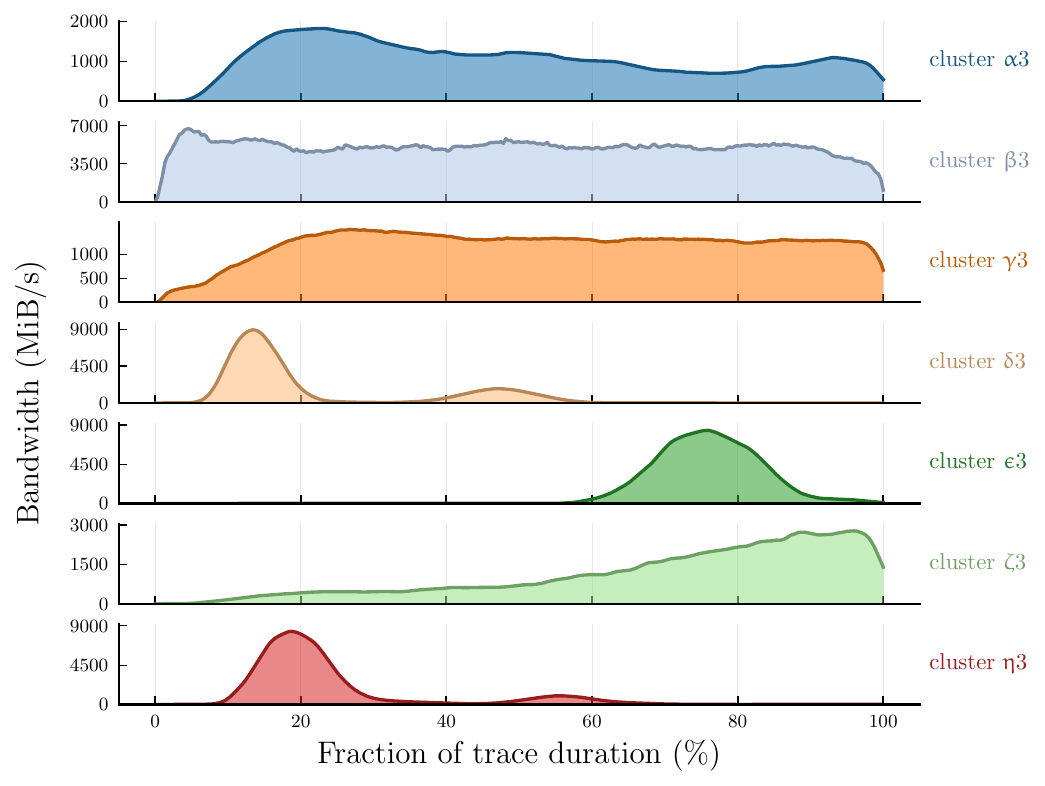}
  \caption{Level \tgen{$3$} clusters produced by \sagetwo. Write.}
  \label{fig:sagetwo-classes-write}
\end{figure}
\end{rronly}

\begin{rronly}
\begin{figure}[t]
  \centering
  \includegraphics[width=\linewidth]{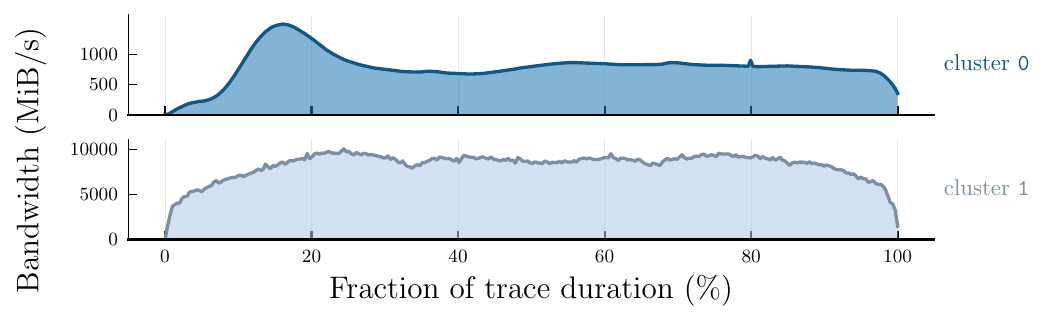}
  \caption{Clusters produced by \tsfresh. Write.}
  \label{fig:tsfresh-classes}
\end{figure}
\end{rronly}

\begin{rronly}
  \todo[inline]{Mention \fig~\ref{fig:per-class} somewhere}
\begin{figure*}[t]
  \centering
  \begin{subfigure}[t]{0.32\textwidth}
    \centering
    \includegraphics[width=\linewidth]{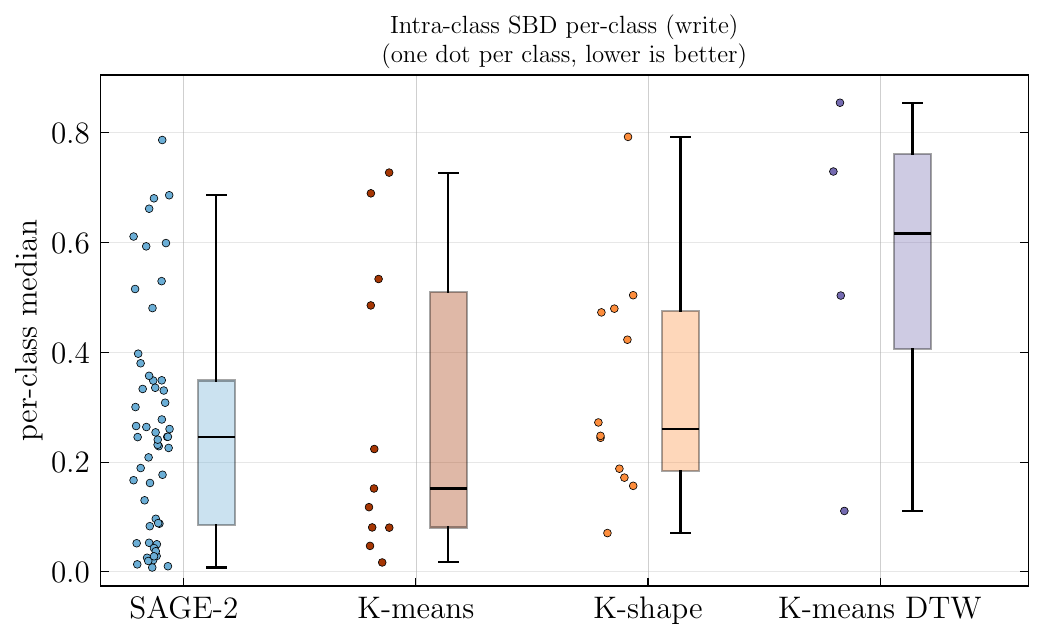}
    \caption{\intra \sbd}
    \label{fig:per-class-intra-sbd}
  \end{subfigure}
  \hfill
  \begin{subfigure}[t]{0.32\textwidth}
    \centering
    \includegraphics[width=\linewidth]{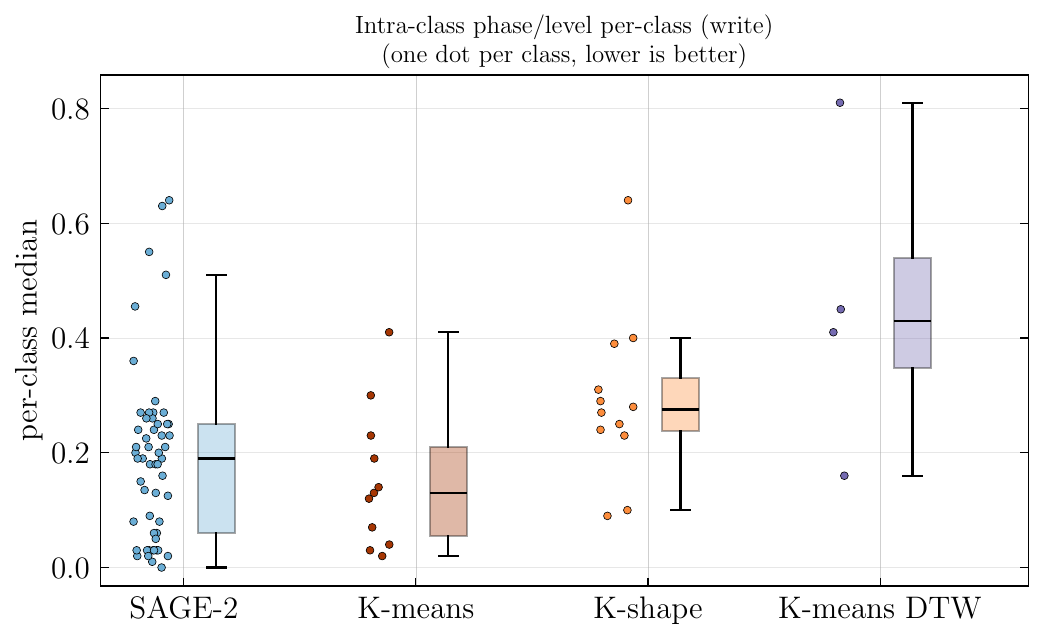}
    \caption{\intra \pld}
    \label{fig:per-class-intra-pld}
  \end{subfigure}
  \hfill
  \begin{subfigure}[t]{0.32\textwidth}
    \centering
    \includegraphics[width=\linewidth]{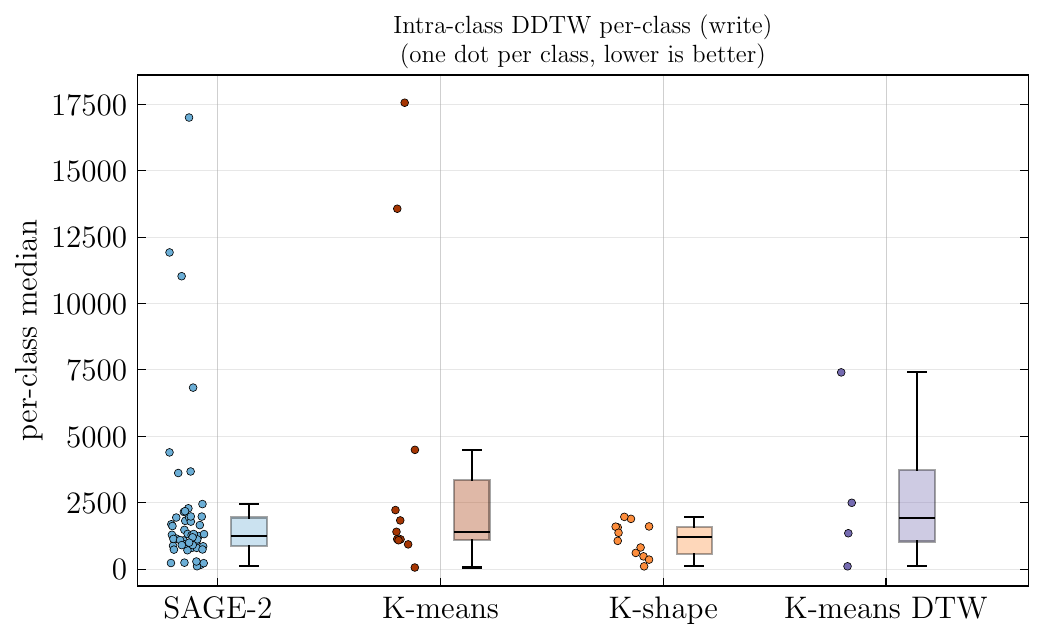}
    \caption{\intra \ddtw}
    \label{fig:per-class-intra-ddtw}
  \end{subfigure}
  \\[0.5em]
  \begin{subfigure}[t]{0.32\textwidth}
    \centering
    \includegraphics[width=\linewidth]{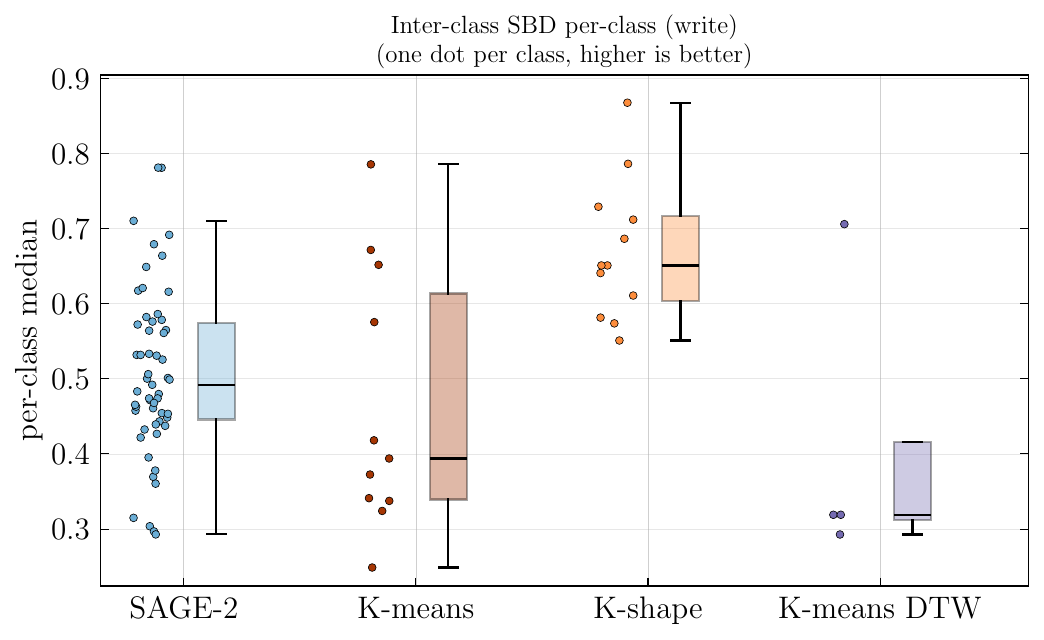}
    \caption{\inter \sbd}
    \label{fig:per-class-inter-sbd}
  \end{subfigure}
  \hfill
  \begin{subfigure}[t]{0.32\textwidth}
    \centering
    \includegraphics[width=\linewidth]{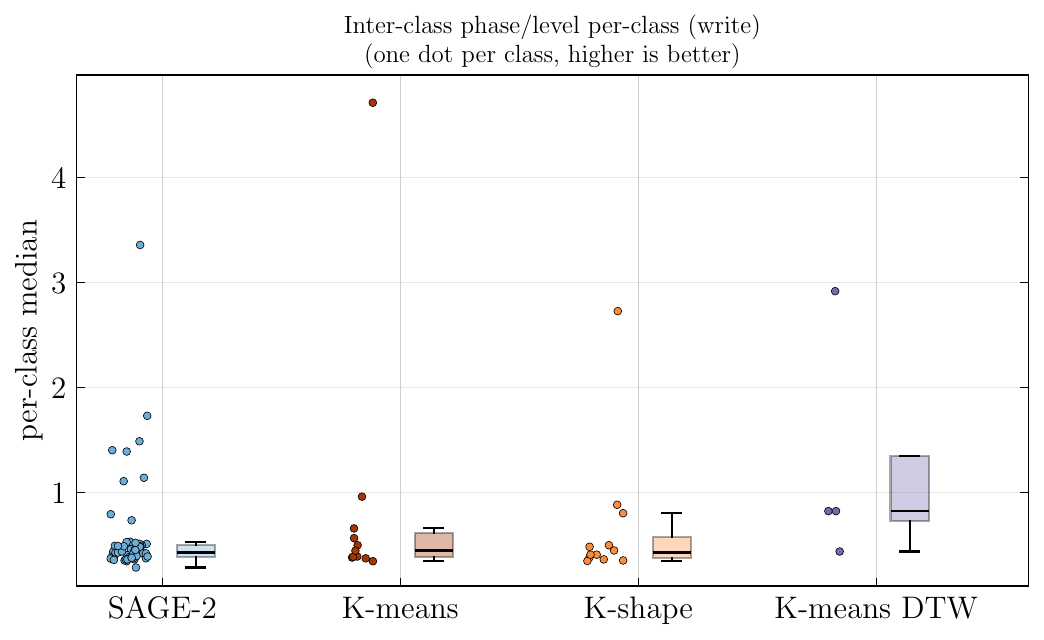}
    \caption{\inter \pld}
    \label{fig:per-class-inter-pld}
  \end{subfigure}
  \hfill
  \begin{subfigure}[t]{0.32\textwidth}
    \centering
    \includegraphics[width=\linewidth]{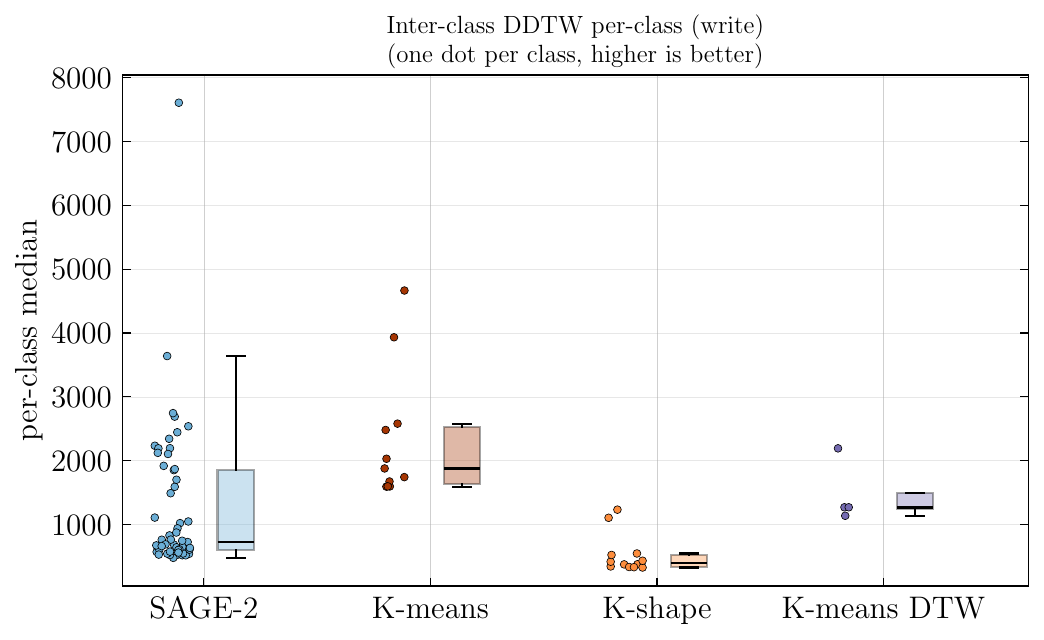}
    \caption{\inter \ddtw}
    \label{fig:per-class-inter-ddtw}
  \end{subfigure}
  \caption{Per-cluster medians. Write. Each dot is one cluster. Boxes span the quartiles. Lower is better for \intra and higher is better for \inter.}
  \label{fig:per-class}
\end{figure*}
\end{rronly}

\subsection{Clustering Time} \label{sec:results_time}

\begin{table}[t]
  \centering
  \caption{Clustering time, in minutes, on \plafrim and \polaris. Includes the elbow search for the number of clusters for the strategies that perform it, namely \sage, \sagetwo, \kmeans, and \tsfresh. OOM: ran out of memory.}
  \label{tab:time}
  \begin{tabular}{lrrrr}
    \toprule
    & \multicolumn{2}{c}{\plafrim} & \multicolumn{2}{c}{\polaris} \\
    \cmidrule(lr){2-3} \cmidrule(lr){4-5}
    Strategy & Read & Write & Read & Write \\
    \midrule
    \random     & 0.2 & 0.1 & 0.5 & 0.2 \\
    \kmeans     & 0.5 & 0.7 & 2.1 & 0.7 \\
    \sagetwo    & 1.8 & 2.1 & 8.3 & 3.0 \\
    \sage       & 4.4 & 4.5 & 16  & 12  \\
    \kshape     & 82  & 229 & 299 & 142 \\
    \tsfresh    & 87  & 70  & OOM & 30  \\
    \kmeansdtw  & 256 & 306 & 309  & 141 \\
    \bottomrule
  \end{tabular}
\end{table}

All times are measured on a single node with two 8-core Intel Xeon E5-2620 CPUs at $2.10$~GHz and $64$~GiB of memory.
\sagetwo is the fastest strategy after \random and \kmeans on every data set (Table~\ref{tab:time}).
\kshape, which matches \sage and \sagetwo in quality on \plafrim, takes $36$ to $109$ times longer.
\kmeansdtw is the slowest strategy overall.

\begin{rronly}
\subsection{Job Length Buckets}\label{sec:job_lengths}

\begin{figure}[t]
  \centering
  \includegraphics[width=\linewidth]{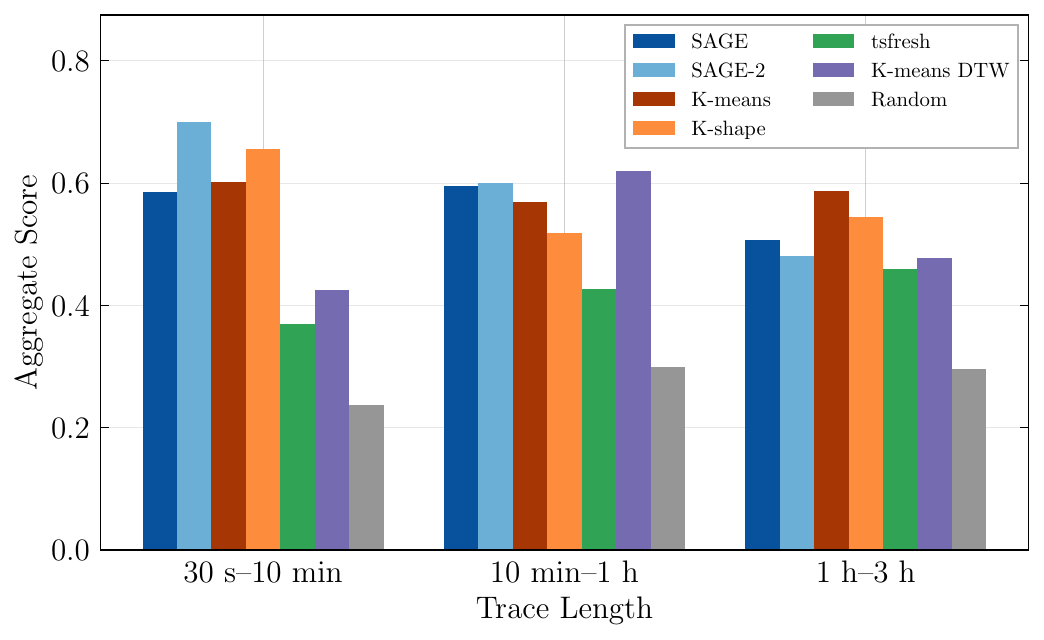}
  \caption{Aggregate score per trace length bucket. Write. Higher is better.
  The score is min-max normalized across the strategies of a bucket, so it compares
  strategies within a bucket but not a strategy across buckets.}
  \label{fig:job-length-aggregate}
\end{figure}
\end{rronly}

\subsection{Does Trace Resolution Impact Results?} \label{sec:results_length}

\begin{figure}[t]
  \centering
  \includegraphics[width=\linewidth]{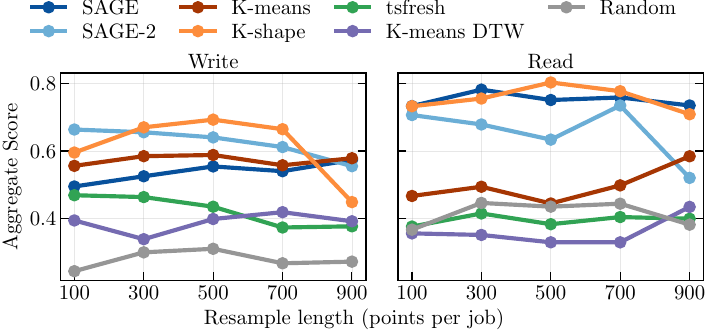}
  \caption{Aggregate score against resample length. Traces of \plafrim. Write (left) and read (right). Higher is better. \rrtext{Min-max normalized across strategies.}}
  \label{fig:resample-sweep}
\end{figure}

\fig~\ref{fig:resample-sweep} shows the impact of the resample length $L$ on the aggregate score on \plafrim.
A larger $L$ increases clustering time. On reads at $L=900$, \sagetwo takes $3.4$ minutes, \kshape $158$ minutes, and \kmeansdtw $887$ minutes, against $1.8$, $82$, and $256$ minutes at the default $L=300$ (Table~\ref{tab:time}).
Since results are stable up to $L=700$ and a larger $L$ only prolongs clustering, there is no reason to use a large value.

\subsection{Summary of Clustering Quality Analysis} \label{sec:results_summary}

Table~\ref{tab:shortcomings} summarizes where each strategy falls short.
\tsfresh assigns nearly every trace to a single cluster, which poorly represents different I/O behaviors. \kmeans is fast but achieves only average cluster balance, \intra distances, and \inter distances. \kshape has a substantially longer clustering time than \sage.
\sage and \sagetwo obtains the best aggregate score on both \plafrim and \polaris, except for \plafrim' writes, where \kshape and \sagetwo are almost tied.
\begin{rronly}
The ``genome'' representation allows the time series to be compressed, making \sage and \sagetwo faster by reducing the amount of data to be clustered.
However, this is \emph{not} its main importance, as compressing them individually would make better approximations and decrease \intra distances. Rather, this step enables our proposal to distinguish traces at a finer granularity than the other strategies.
\end{rronly}
These metrics, however, only measure how well the clusters represent the traces. 
In Section~\ref{sec:simulation}, we evaluate whether good scores translate into better decisions in an I/O scheduling use case. 

\begin{table}[t]
  \centering
  \caption{Summary of the clustering quality analysis on \plafrim writes: \good good, \mixed mixed, \bad poor.}
  \label{tab:shortcomings}
  \setlength{\tabcolsep}{3.5pt}
  \begin{tabular}{@{}lcc@{\hspace{10pt}}cccc@{}}
    \toprule
            & \sage  & \sagetwo & \kmeans & \kshape & \tsfresh & \kmeansdtw \\
    \midrule
    Time    & \good  & \good  & \good  & \bad   & \bad   & \bad   \\
    \intra  & \good  & \good  & \mixed & \good  & \mixed & \bad   \\
    \inter  & \mixed & \mixed & \mixed & \mixed & \mixed & \mixed \\
    Balance & \good  & \good  & \mixed & \good  & \bad   & \mixed \\
    \bottomrule
  \end{tabular}
\end{table}

\section{Use Case: OST Scheduling from Cluster Centroids}\label{sec:simulation} \label{sec:use_case_scheduling}

As discussed in Section~\ref{sec:introduction}, I/O scheduling techniques need to know when jobs do I/O in order to mitigate interference.
One such technique is the scheduling of object storage targets (OSTs)~\cite{boito:hal-05662894}.
A PFS stores data on several OSTs, and each I/O request accesses one or more of them.
If we consider only data access (and ignore metadata and network effects), jobs that use different OSTs do not interfere, because they share no I/O resource.
The problem is then to decide which OSTs each job uses, that is, which jobs share them, so as to minimize the competition for bandwidth~\cite{bandet:hal-04394004}.

In practice, however, the exact time series of a job is only known after its execution.
As argued in Section~\ref{sec:introduction}, predicting the exact temporal behavior of a job is hard, whereas identifying its cluster is a more realistic goal.
We therefore consider a scheduler that only knows the cluster of each job, and uses the cluster centroids for placement decisions.
Our assumption is that a better clustering produces centroids that represent jobs well, and therefore gives better information for placement. 


In this section, we evaluate the clustering strategies on this problem.
To do so, we use a simulator built on top of SimGrid~\cite{simgrid} (Section~\ref{sec:sim-model}), in which a placement algorithm (Section~\ref{sec:sim-placement}) selects one OST for each job. 

\subsection{Simulation Model} \label{sec:sim-model}

We rely on SimGrid $4.1$, a mature framework for simulating distributed platforms whose resource-sharing models have been validated against real systems~\cite{simgrid}, and which has already been used to evaluate I/O scheduling techniques~\cite{iosets}.

We built a simulation where each OST is a disk (we use 4 OSTs) whose bandwidth is equally shared by competing jobs. Jobs are represented by a sequence of compute and I/O phases, and a placement algorithm decides, once per job (for simplicity), the OST it should use. 
To build the job phases, we take traces from the \plafrim data set. From each trace, all time spent at bandwidth lower than $1$~MiB/s becomes compute time (duration does not depend on the schedule), and time spent on I/O becomes an I/O phase defined by the amount of data that was moved (so its duration will depend on the obtained bandwidth).
\begin{rronly}
\fig~\ref{fig:sim-phases} illustrates the process of turning traces into jobs for the simulation.
\begin{figure}[t]
  \centering
  \begin{tikzpicture}[x=0.6cm, y=1cm, font=\scriptsize]
    \node[anchor=east, align=right] at (-0.2,0.55) {trace\\[-1pt](MiB/s)};
    \foreach \v [count=\i from 0] in {0.2,0.5,400,450,0.4,0,0.8,300} {
      \pgfmathsetmacro{\h}{max(\v/450*1.1,0.04)}
      \pgfmathparse{int(\v>=1)}
      \ifnum\pgfmathresult=1
        \fill[blue!55] (\i+0.12,0) rectangle (\i+0.88,\h);
      \else
        \fill[black!35] (\i+0.12,0) rectangle (\i+0.88,\h);
      \fi
      \node[font=\tiny, black!60] at (\i+0.5,-0.15) {\v};
    }
    \draw[orange!80!black, dashed, line width=0.6pt] (-0.05,0.09) -- (8.05,0.09);
    \node[anchor=west, orange!80!black, align=left, font=\tiny] at (8.15,0.12) {$1$~MiB/s\\[-2pt](not to scale)};

    \node[anchor=east] at (-0.2,-0.65) {phases};
    \fill[black!20] (0,-0.8) rectangle (2,-0.5);
    \fill[blue!55] (2,-0.8) rectangle (4,-0.5);
    \fill[black!20] (4,-0.8) rectangle (7,-0.5);
    \fill[blue!55] (7,-0.8) rectangle (8,-0.5);
    \node at (1,-0.65) {cpu 2\,s};
    \node at (3,-0.65) {850\,MiB};
    \node at (5.5,-0.65) {cpu 3\,s};
    \node[font=\tiny] at (7.5,-0.65) {300\,MiB};

    \node[anchor=east, align=right] at (-0.2,-1.3) {alone on\\[-1pt]an OST};
    \fill[black!20] (0,-1.45) rectangle (2,-1.15);
    \fill[blue!55] (2,-1.45) rectangle (5.4,-1.15);
    \fill[black!20] (5.4,-1.45) rectangle (8.4,-1.15);
    \fill[blue!55] (8.4,-1.45) rectangle (9.6,-1.15);
    \node at (1,-1.3) {cpu 2\,s};
    \node at (3.7,-1.3) {3.4\,s};
    \node at (6.9,-1.3) {cpu 3\,s};
    \node[font=\tiny] at (9,-1.3) {1.2\,s};
    \draw[-{Latex[length=3pt]}, black!50] (0,-1.6) -- (10,-1.6) node[right] {time};
  \end{tikzpicture}
  \caption{From a trace to phases.
  Samples of at least $1$~MiB/s are I/O (blue), the others computation (grey), and each run of samples becomes a phase.
  A compute phase keeps its duration, while an I/O phase keeps its volume, so its duration depends on the bandwidth it gets: alone on an OST of $250$~MiB/s, the $850$~MiB take $3.4$~s.}
  \label{fig:sim-phases}
\end{figure}

\end{rronly}

To measure the impact of interference, we use the \emph{I/O stretch} of a job $j$, defined as 
$T_j/T^{isol}_{j}$, 
where $T_j$ is the total time spent by the I/O phases of $j$ during the simulation, and $T^{isol}_{j}$ is the time they would take in isolation. It therefore measures how many times slower the I/O of a job was because it competed with others. We take the geometric mean of the I/O stretch of all jobs in an execution.

\subsection{Placement Algorithm} \label{sec:sim-placement}

Our greedy placement algorithm relies on the notion of \emph{I/O overlap}.
We compare all pairs of jobs $a$ and $b$, using their expected time series in isolation and their start times, and measure the time $t_{ov}$ during which they would do I/O at the same time. The expected time series comes from the cluster centroid, re-scaled to match the job duration. We compute I/O overlap as the mean of the proportion of time spent competing for I/O:
\begin{equation}\label{eq:overlap}
  w(a,b) = \frac{1}{2}\left(\frac{t_{ov}}{T^{isol}_{a}} + \frac{t_{ov}}{T^{isol}_{b}}\right).
\end{equation}
It goes from $0$, when the two jobs never do I/O at the same time, to $1$, when they always 
\begin{cponly}
  do.
\end{cponly}
\begin{rronly}
do, as in the example of \fig~\ref{fig:sim-overlap}.
  
\begin{figure}[t]
  \centering
  \begin{tikzpicture}[x=0.52cm, y=1cm, font=\scriptsize]
    \node[anchor=west, font=\scriptsize\bfseries] at (-2.2,2.45) {(a) real traces};
    \node[anchor=east] at (-0.15,1.9) {job $a$};
    \node[anchor=east] at (-0.15,1.3) {job $b$};
    \fill[black!10] (2,1.08) rectangle (3,2.12);
    \fill[black!10] (6,1.08) rectangle (7,2.12);
    \fill[blue!55] (1,1.75) rectangle (3,2.05);
    \fill[blue!55] (6,1.75) rectangle (8,2.05);
    \fill[orange!80] (2,1.15) rectangle (7,1.45);
    \draw[-{Latex[length=3pt]}, black!50] (0,0.95) -- (9.2,0.95) node[right] {time};
    \foreach \x in {0,...,8} \node[below, font=\tiny, black!60] at (\x,0.95) {\x};
    \node[anchor=west, align=left] at (9.9,1.6) {$t_{io,a}=4$, $t_{io,b}=5$\\$t_{ov}=2$, $w=0.45$};

    \node[anchor=west, font=\scriptsize\bfseries] at (-2.2,0.2) {(b) same centroid};
    \node[anchor=east] at (-0.15,-0.35) {job $a$};
    \node[anchor=east] at (-0.15,-0.95) {job $b$};
    \fill[black!10] (2.8,-1.17) rectangle (3.15,-0.13);
    \fill[black!25] (0,-0.5) rectangle (4,-0.2);
    \fill[blue!55] (0.8,-0.5) rectangle (1.4,-0.2);
    \fill[blue!55] (2.8,-0.5) rectangle (3.6,-0.2);
    \fill[black!25] (0,-1.1) rectangle (9,-0.8);
    \fill[orange!80] (1.8,-1.1) rectangle (3.15,-0.8);
    \fill[orange!80] (6.3,-1.1) rectangle (8.1,-0.8);
    \draw[-{Latex[length=3pt]}, black!50] (0,-1.3) -- (9.2,-1.3) node[right] {time};
    \node[anchor=west, align=left] at (9.9,-0.65) {same length: $w=1$\\stretched: $w=0.18$};
  \end{tikzpicture}
  \caption{I/O overlap of two jobs, with I/O phases in color and overlaps shaded.
  (a) From the real traces, with \autoref{eq:overlap}.
  (b) From the centroid of their cluster, which does I/O between $20$ and $35\%$ and between $70$ and $90\%$ of the run.
  With a single length, the two jobs would have the same timeline and $w=1$.
  Stretched to the duration of each job (grey bars), they only overlap briefly.
  For readability, both jobs start at time $0$.}
  \label{fig:sim-overlap}
\end{figure}
\end{rronly}

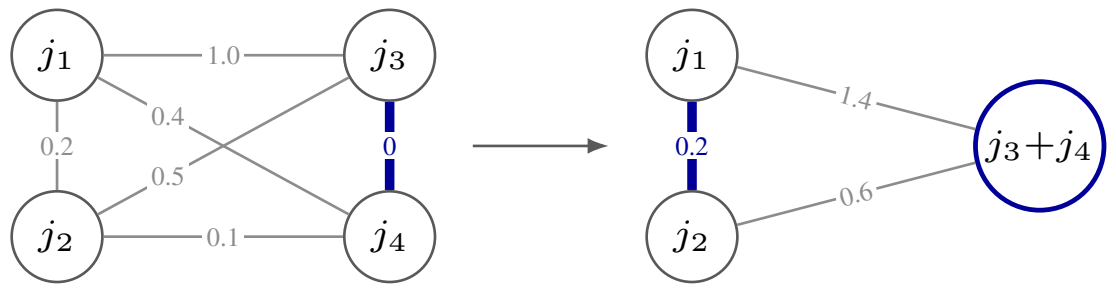
\begin{figure}[t]
  \centering
  \begin{tikzpicture}[font=\scriptsize,
      gnode/.style={circle, draw=black!65, fill=white, line width=0.5pt, minimum size=0.6cm, inner sep=0pt},
      gedge/.style={line width=0.5pt, black!45},
      glabel/.style={font=\tiny, fill=white, inner sep=0.8pt},
      gnext/.style={line width=1.8pt, blue!60!black}]
    \begin{scope}
      \node[gnode] (a1) at (0,1.2) {$j_1$};
      \node[gnode] (a3) at (2.2,1.2) {$j_3$};
      \node[gnode] (a2) at (0,0) {$j_2$};
      \node[gnode] (a4) at (2.2,0) {$j_4$};
      \draw[gedge] (a1) -- node[glabel] {1.0} (a3);
      \draw[gedge] (a1) -- node[glabel] {0.2} (a2);
      \draw[gedge] (a2) -- node[glabel] {0.1} (a4);
      \draw[gedge] (a1) -- node[glabel, pos=0.28] {0.4} (a4);
      \draw[gedge] (a2) -- node[glabel, pos=0.28] {0.5} (a3);
      \draw[gnext] (a3) -- node[glabel, text=blue!60!black] {0} (a4);
    \end{scope}
    \draw[wfarr] (2.75,0.6) -- (3.65,0.6);
    \begin{scope}[xshift=4.2cm]
      \node[gnode] (b1) at (0,1.2) {$j_1$};
      \node[gnode] (b2) at (0,0) {$j_2$};
      \node[gnode, minimum size=0.85cm, draw=blue!60!black, line width=0.9pt] (b34) at (2.3,0.6) {$j_3{+}j_4$};
      \draw[gedge] (b1) -- node[glabel, sloped] {1.4} (b34);
      \draw[gedge] (b2) -- node[glabel, sloped] {0.6} (b34);
      \draw[gnext] (b1) -- node[glabel, text=blue!60!black] {0.2} (b2);
    \end{scope}
  \end{tikzpicture}
  \caption{Greedy merging on four jobs and $<4$ OSTs.
  The edges of the merged node are the sums of the edges they replace ($1.4 = 1.0 + 0.4$).}
  \label{fig:sim-merge}
\end{figure}

\paragraph{Step 1: greedy merging}
The algorithm builds a complete graph where nodes are jobs, and edges are weighted by their I/O overlap.
  While the number of nodes is greater than the number of OSTs, it merges the two nodes joined by the lightest edge (ties are broken by selecting the nodes with the fewest jobs). The collapsed edges are updated with the sum of their weights.
This process is illustrated in \fig~\ref{fig:sim-merge}.

\paragraph{Step 2: local search}
At the end of step 1, each node contains the jobs that will be placed on the same OST. A local search then improves the placement by considering all possible moves or swaps until no change can decrease the overlap between co-located jobs.
  
\begin{rronly}
It considers two kinds of changes: moving one job to another OST, and swapping two jobs placed on different OSTs.
Let $ov(i,o)$ be the sum of the I/O overlaps of job $i$ with the jobs placed on OST $o$.
When job $i$ moves from OST $o_1$ to OST $o_2$, it stops colliding with the jobs of $o_1$ and starts colliding with the jobs of $o_2$, so the total overlap changes by
\begin{equation}\label{eq:move}
  \Delta_{move}(i, o_1 \rightarrow o_2) = ov(i,o_2) - ov(i,o_1).
\end{equation}
A swap of job $i$ on $o_1$ with job $j$ on $o_2$ is equivalent to two such moves, of $i$ to $o_2$ and of $j$ to $o_1$.
Each of these moves, however, is computed as if the other job stayed on its OST, so the overlap $w(i,j)$ between $i$ and $j$ is counted once in each of them, although $i$ and $j$ never share an OST, neither before nor after the swap.
The change of a swap is therefore
\begin{equation}\label{eq:swap}
  \Delta_{swap} = \Delta_{move}(i, o_1 \rightarrow o_2) + \Delta_{move}(j, o_2 \rightarrow o_1) - 2\,w(i,j).
\end{equation}
At each round, the local search applies the move or swap with the most negative change, and it stops when no change is negative, that is, when no single move or swap can reduce the total overlap anymore.
\end{rronly}

\subsection{Workloads and Settings}\label{sec:sim-workloads}

For a simulation time (or horizon) $H$, and $O$ OSTs, we 
define the \emph{storage load} of a workload of $n$ jobs as
\begin{equation}
  \rho = \frac{\sum_{i=1}^{n} T^{isol}_{i}}{O \cdot H},
\end{equation}
that is, the I/O time requested by all jobs divided by the I/O time available over $H$.
With 
$\rho > 1$, the storage is overloaded and jobs must share OSTs while doing I/O.

We build workloads with \plafrim read traces lasting between $10$ and $60$ minutes ($H=60$). We randomly draw workloads of $n$ jobs until their load is within $0.02$ of the target $\rho$, then we draw start times at random so that all jobs start and end within the simulated time.
\begin{rronly}
On a workload of $40$ jobs with $\rho = 1$, $37\%$ of the pairs of jobs overlap, against $81\%$ if every job started at time $0$.
\end{rronly}
We build workloads of $40$ and $80$ jobs, with loads $\rho$ of $0.5$, $1$, $1.5$, $2$, $2.5$ and $3$, and $10$ random workloads per load.
\begin{rronly}
Giving a total of $60$ workloads per number of jobs.
\end{rronly}

We then simulate each workload with the placement algorithm of Section~\ref{sec:sim-placement}, using the output of each clustering strategy. We also add three baselines: round-robin and random placement, which require no information about jobs, and \clairvoyant, which uses the actual time series of jobs instead of clustering information.
The latter is our ``best-case'' reference. 
Among the clustering strategies, we show the placements obtained with \sage, \sagetwo, \kmeans (the most widely used clustering technique), \kshape (the best baseline in the clustering evaluation from Section~\ref{sec:evaluation}), and \kmeansdtw.
\begin{cponly}
\textcolor{blue}{The results for all strategies will be available in a research report after publication.}
\end{cponly}
\begin{rronly}
\added{\fig~\ref{fig:sim-heatmaps-all-read} and \fig~\ref{fig:sim-heatmaps-all-write} show the results of all strategies.}
\end{rronly}

\subsection{Results}\label{sec:sim-results}

\begin{figure}[t]
  \centering
  \includegraphics[width=\columnwidth]{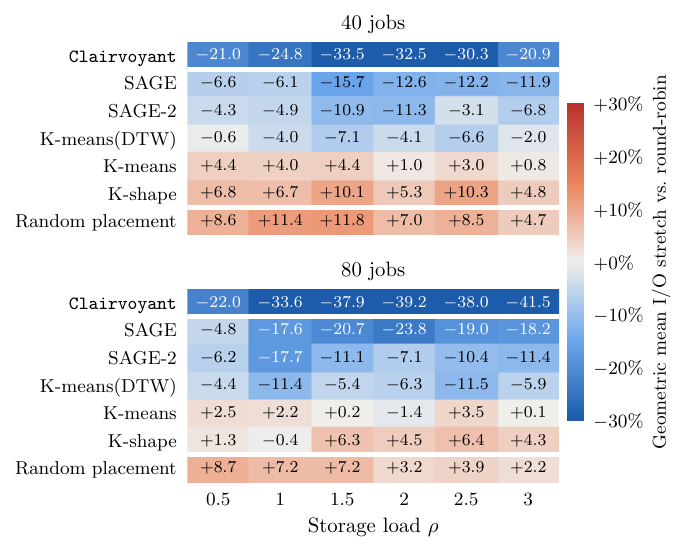}
  \caption{Difference in I/O stretch (\%) compared to round-robin. \textbf{Lower is better}, with negative meaning an improvement. Read traces of \plafrim.}
  \label{fig:sim-heatmaps}
\end{figure}

\fig~\ref{fig:sim-heatmaps} shows our results. For each workload, we report the difference in I/O stretch against round-robin, so that a negative value means an improvement. We can see that taking jobs' temporal behaviors into consideration for placement (the first row, using the precise information) improves on round-robin in every scenario, by up to $42\%$. The potential for improvement grows with the number of jobs.

\sage, which has the best aggregate score on the read traces (\fig~\ref{fig:aggregate}), also improves on round-robin in every scenario, with the largest average reductions of all clustering strategies: it decreases the I/O stretch by up to $24\%$. It is followed by \sagetwo and \kmeansdtw, which decrease it by up to $18\%$ and $12\%$, respectively. The clustering of \kmeansdtw, however, is about $58\times$ slower than that of \sage on the read traces (Table~\ref{tab:time}).
\kmeans and \kshape, on the other hand, lead to placements worse than round-robin in most scenarios.
\begin{cponly}
The experiments with write traces have similar results (some are shown in \fig~\ref{fig:motivation-simulation}) and \textcolor{blue}{will be available in a research report after publication.}
\end{cponly}
\begin{rronly}
\added{\fig~\ref{fig:sim-heatmaps-write} shows the complete results on write traces. \kmeans benefits more from write traces than from read traces, and improves on round-robin in every scenario, although less than the genome strategies.}
\added{Write traces leave more room for placement, as \clairvoyant also gains more on them.
On write traces, a few jobs do most of the I/O, while most jobs write very little.
With $80$ jobs, for instance, the $8$ heaviest jobs do $69\%$ of the I/O time when writing, against $57\%$ when reading.
Placing these few jobs on different OSTs therefore removes most of the contention.
This also helps \kmeans, whose centroids predict the I/O overlap much better for writes than for reads.}

\begin{figure}[t]
  \centering
  \includegraphics[width=\columnwidth]{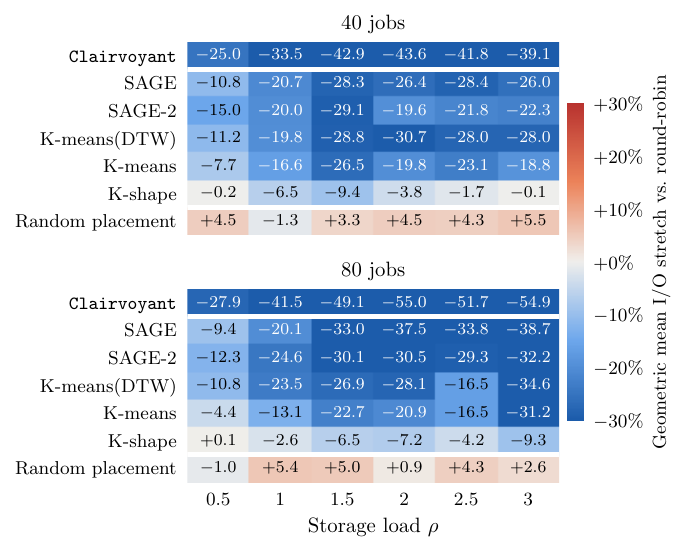}
  \caption{\added{Same as \fig~\ref{fig:sim-heatmaps}, on write traces.}}
  \label{fig:sim-heatmaps-write}
\end{figure}

\begin{figure*}[t]
  \centering
  \includegraphics[width=\textwidth]{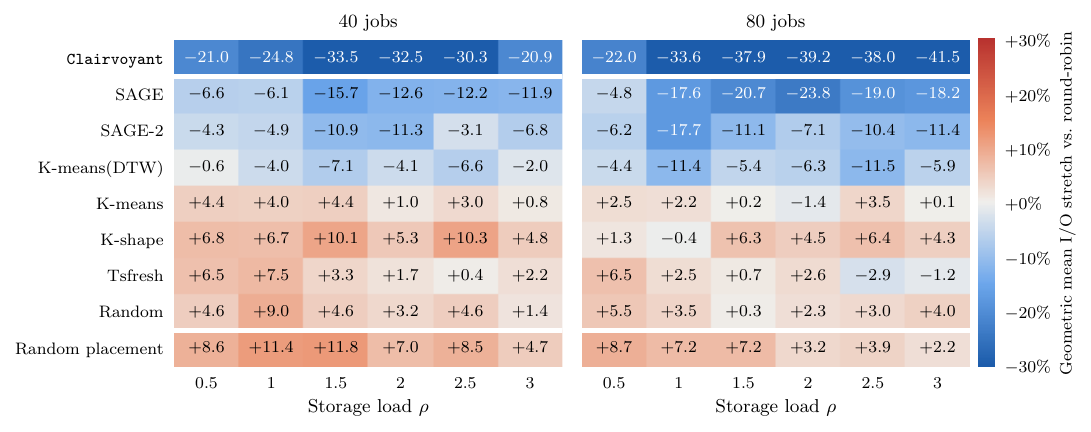}
  \caption{\added{Geometric mean I/O stretch difference (\%) compared to round-robin on $4$ OSTs for all clustering strategies, read traces. Lower is better.}}
  \label{fig:sim-heatmaps-all-read}
\end{figure*}

\begin{figure*}[t]
  \centering
  \includegraphics[width=\textwidth]{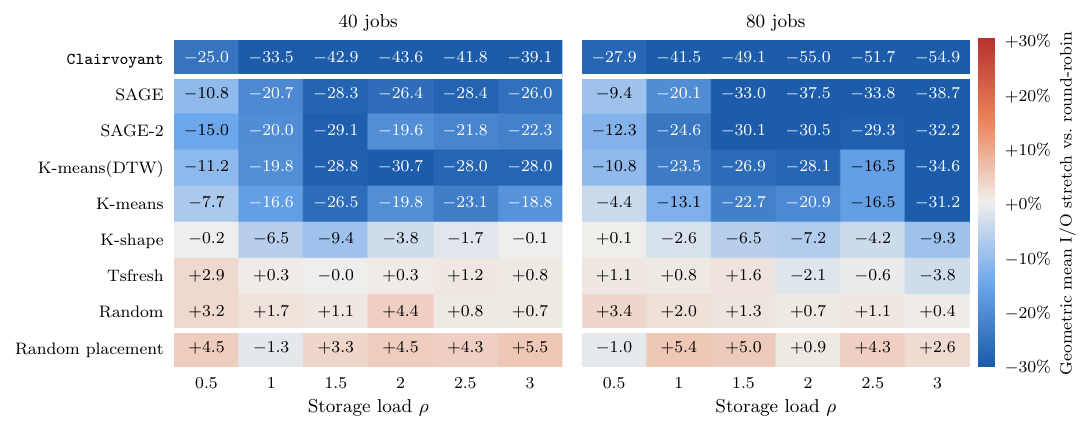}
  \caption{\added{Same as \fig~\ref{fig:sim-heatmaps-all-read}, on write traces.}}
  \label{fig:sim-heatmaps-all-write}
\end{figure*}
\end{rronly}
\begin{rronly}
\added{Only \clairvoyant does better than \sage,} but it requires the traces of the jobs, which are only known once the jobs have finished and are therefore not available in practice.
The clusters, on the other hand, can be built beforehand from past traces, and then used to place the future jobs of the machine.
\added{\sagetwo also improves on round-robin in every configuration, but with smaller gains.
\kmeans, \kshape, \tsfresh and the \random clusters are worse than round-robin on average, while \kmeansdtw improves on it.
The centroids of the strategies that do worse than round-robin may average members that do I/O at different moments of their run, so that they overlap with most other jobs.}
The placement then acts on overlaps that do not match the real ones, which can be worse than not using any information.
\luan{This explanation comes from the shape of the centroids, we should check it on the \kshape, \kmeansdtw and \tsfresh centroids of this range of durations.}
\end{rronly}

\begin{rronly}
The gains grow with the number of jobs, for \added{\clairvoyant as for \sage}.
At the same load, more jobs means lighter jobs: a job spends on average \added{$35\%$ of its run in I/O with $40$ jobs and $21\%$} with $80$ jobs.
Lighter jobs leave gaps between their I/O phases, so fewer pairs of jobs ever do I/O at the same time, and many pairs can share an OST at no cost.
With $10$ or $20$ jobs per OST, the placement also has many more combinations to choose from.
Round-robin cannot use these gaps, while a placement that knows the I/O overlap avoids more of the contention.
The placement from \added{\sage} centroids follows the same trend.
\end{rronly}

\section{Limitations}\label{sec:limitation}




Our approach proposes an accurate I/O time series classification that addresses the unique I/O patterns. Specifically, SAGE specializes meta-parameters (e.g., number of symbols $L$ to address various durations, size of the window) and performs a dedicated encoding per system. However, transferring our workflow to different systems or time series signals (e.g., energy consumption, performance counters) may require adjustments to the meta-parameters or the encoding to capture their properties. Studying these adjustments is interesting to identify common patterns or differences from data across various domains of research. 
Moreover, the resampling to a common length $L$ hides the frequency of I/O phases for periodic applications. To avoid losing this information, we would recommend using \sage in addition to some periodicity characterization technique such as FTIO~\cite{10579160}.

Finally, 
we relied on simulation to provide a proof of concept for the clustering-based OST scheduling use case. Our focus for future work is to implement this full workflow, including the prediction of classes, into a real system.

\section{Conclusion}\label{sec:conclusion}

We presented \sage and \sagetwo, two techniques to cluster HPC jobs from their I/O activity over time. 
Our approach 
cuts each trace into short windows, learns an alphabet of shapes and amplitude levels from these windows, and translates each trace into a sequence of symbols before clustering it.
We also proposed \pld, a metric that accounts for both when a job does I/O and at which bandwidth, tackling weaknesses of existing metrics. All source code required to apply our approach to new data sets is publicly available~[\textcolor{blue}{once the paper is published}]. Finally, we release a data set with the I/O traces of \numberofpublishedtraces jobs to support further work in the field.

We evaluated our approach
on \totalnumberoftraces traces from \numberofdatasets systems. It obtains the best aggregate score in three of the four cases, while \kshape, the closest baseline in quality, takes $36$ to $109$ times longer to produce a clustering.

Clustering metrics are useful to compare strategies, but they are not enough to judge whether the clusters are useful.
We therefore evaluated the clusters on a downstream task, the placement of jobs on the OSTs of a PFS, where the scheduler only knows the centroid of the cluster of each job.
In this task, using our clusters, we \improvementonstretch.
The ranking of the strategies, however, does not follow the metrics exactly.
\kshape, whose aggregate score is close to ours, leads to placements worse than round-robin on read traces, whereas \kmeansdtw, which, along with \tsfresh, has the lowest aggregate score, gives good placements on write traces.
On write traces, \sagetwo has the best aggregate score, but \sage gives slightly better placements.
The downstream task is therefore the best judge of a clustering.
Under this criterion, \sage and \sagetwo remain the best strategies on read traces and among the best on write traces.

\section*{Acknowledgment}
As part of the "France 2030" initiative, this work has benefited from a State grant managed by the French National Research Agency (Agence Nationale de la Recherche) attributed to the Exa-DoST project of the NumPEx PEPR program, reference: ANR-22-EXNU-0004.

Experiments presented in this paper were carried out using Grid5000, one of the precursor infrastructures of the SLICES-FR testbed, operated by a scientific interest group including 14 major French academic institutes and Universities (see https://slices-fr.eu/).

Experiments presented in this paper were carried out using the PlaFRIM experimental testbed, supported by Inria, CNRS (LABRI and IMB), Université de Bordeaux, Bordeaux INP and Conseil Régional d’Aquitaine (see https://www.plafrim.fr). 

This data was generated from resources of the Argonne Leadership Computing Facility, which is a DOE Office of Science User Facility supported under Contract DE-AC02-06CH11357.

\bibliographystyle{IEEEtran}
\bibliography{ref}

@inproceedings{pavan2019unsupervised,
  title={An unsupervised learning approach for I/O behavior characterization},
  author={Pavan, Pablo J and Bez, Jean Luca and Serpa, Matheus S and Boito, Francieli Zanon and Navaux, Philippe OA},
  booktitle={2019 31st International Symposium on Computer Architecture and High Performance Computing (SBAC-PAD)},
  pages={33--40},
  year={2019},
  organization={IEEE}
}

@article{new_shabe_based_metric,
title = {A new shape-based clustering algorithm for time series},
journal = {Information Sciences},
volume = {609},
pages = {411-428},
year = {2022},
issn = {0020-0255},
doi = {https://doi.org/10.1016/j.ins.2022.07.105},
url = {https://www.sciencedirect.com/science/article/pii/S0020025522007940},
author = {Yucheng Li and Derong Shen and Tiezheng Nie and Yue Kou}
}

@article{sbd_funding_paper,
author = {Paparrizos, John and Gravano, Luis},
title = {Fast and Accurate Time-Series Clustering},
year = {2017},
issue_date = {June 2017},
publisher = {Association for Computing Machinery},
address = {New York, NY, USA},
volume = {42},
number = {2},
issn = {0362-5915},
url = {https://doi.org/10.1145/3044711},
doi = {10.1145/3044711},
journal = {ACM Trans. Database Syst.},
month = jun,
articleno = {8},
numpages = {49}
}

@article{shannon1948,
  title={A Mathematical Theory of Communication},
  author={Shannon, Claude E},
  journal={Bell System Technical Journal},
  volume={27},
  number={3},
  pages={379–423},
  year={1948}
}

@article{kshape_funding_paper,
author = {Paparrizos, John and Gravano, Luis},
title = {k-Shape: Efficient and Accurate Clustering of Time Series},
year = {2016},
issue_date = {March 2016},
publisher = {Association for Computing Machinery},
address = {New York, NY, USA},
volume = {45},
number = {1},
issn = {0163-5808},
url = {https://doi.org/10.1145/2949741.2949758},
doi = {10.1145/2949741.2949758},
journal = {SIGMOD Rec.},
month = jun,
pages = {69–76},
numpages = {8}
}

@inproceedings{metis,
author = {Youssef, Karim and Neuwirth, Sarah and Rajesh, Neeraj and Devarajan, Hariharan},
title = {Metis: Agentic Knowledge Synthesis for Explainable I/O Performance in HPC Systems},
year = {2026},
isbn = {9798400726408},
publisher = {Association for Computing Machinery},
address = {New York, NY, USA},
url = {https://doi.org/10.1145/3806645.3816129},
doi = {10.1145/3806645.3816129},
booktitle = {Proceedings of the 35th International Symposium on High-Performance Parallel and Distributed Computing},
pages = {817–826},
numpages = {10},
location = {
},
series = {HPDC '26}
}

@inproceedings{user_based_io_profiling,
author = {Yazdani, Ahmad Hossein and Paul, Arnab K. and Karimi, Ahmad Maroof and Wang, Feiyi and Butt, Ali},
title = {User-based I/O Profiling for Leadership Scale HPC Workloads},
year = {2025},
isbn = {9798400710629},
publisher = {Association for Computing Machinery},
address = {New York, NY, USA},
url = {https://doi.org/10.1145/3700838.3700865},
doi = {10.1145/3700838.3700865},
booktitle = {Proceedings of the 26th International Conference on Distributed Computing and Networking},
pages = {181–190},
numpages = {10},
location = {
},
series = {ICDCN '25}
}

@inproceedings{10.1145/3391812.3396270,
author = {Bang, Jiwoo and Kim, Chungyong and Wu, Kesheng and Sim, Alex and Byna, Suren and Kim, Sunggon and Eom, Hyeonsang},
title = {HPC Workload Characterization Using Feature Selection and Clustering},
year = {2020},
isbn = {9781450379809},
publisher = {Association for Computing Machinery},
address = {New York, NY, USA},
url = {https://doi.org/10.1145/3391812.3396270},
doi = {10.1145/3391812.3396270},
booktitle = {Proceedings of the 3rd International Workshop on Systems and Network Telemetry and Analytics},
pages = {33–40},
numpages = {8},
location = {Stockholm, Sweden},
series = {SNTA '20}
}

@article{Halawa_2020,
   title={Unsupervised KPIs-Based Clustering of Jobs in HPC Data Centers},
   volume={20},
   ISSN={1424-8220},
   url={http://dx.doi.org/10.3390/s20154111},
   DOI={10.3390/s20154111},
   number={15},
   journal={Sensors},
   publisher={MDPI AG},
   author={Halawa, Mohamed S. and Díaz Redondo, Rebeca P. and Fernández Vilas, Ana},
   year={2020},
   month=jul, 
   pages={4111} }

@unpublished{ahmadzada:hal-05658205,
  TITLE = {{Sufficiency in Data Centers: Energy-Aware Resource Recommendation System}},
  AUTHOR = {Ahmadzada, Eyvaz and Stolf, Patricia and Pierson, Jean-Marc and Lefevre, Laurent},
  URL = {https://hal.science/hal-05658205},
  NOTE = {This document is the preprint version of the paper. The accepted manuscript and publisher's version are currently subject to an embargo imposed by the publisher. They will be made available after the embargo period ends, in compliance with the publisher's open access policy.},
  YEAR = {2026},
  BOOKTITLE={2026 Europar}, 
  MONTH = Jun,
  HAL_ID = {hal-05658205},
  HAL_VERSION = {v1},
}

@InProceedings{Evalix,
author="Emeras, Joseph
and Varrette, S{\'e}bastien
and Guzek, Mateusz
and Bouvry, Pascal",
editor="Desai, Narayan
and Cirne, Walfredo",
title="Evalix: Classification and Prediction of Job Resource Consumption on HPC Platforms",
booktitle="Job Scheduling Strategies for Parallel Processing",
year="2017",
publisher="Springer International Publishing",
address="Cham",
pages="102--122",
isbn="978-3-319-61756-5"
}

@INPROCEEDINGS{IOscout,
  author={Li, Yuqi and Xiao, Liquan and Feng, Jinghua and Zhang, Jian and Zheng, Gang and Yuan, Yuan},
  booktitle={2023 IEEE 3rd International Conference on Computer Communication and Artificial Intelligence (CCAI)}, 
  title={IOScout: an I/O Characteristics Prediction Method for the Supercomputer Jobs}, 
  year={2023},
  volume={},
  number={},
  pages={205-210},
  doi={10.1109/CCAI57533.2023.10201270}}

@inproceedings{bang2021workflow,
author = {Kunkel, Julian and Betke, Eugen},
title = {Toward a Workflow for Identifying Jobs with Similar I/O Behavior Utilizing Time Series Analysis},
year = {2021},
isbn = {978-3-030-90538-5},
publisher = {Springer-Verlag},
address = {Berlin, Heidelberg},
url = {https://doi.org/10.1007/978-3-030-90539-2_10},
doi = {10.1007/978-3-030-90539-2_10},
booktitle = {High Performance Computing: ISC High Performance Digital 2021 International Workshops, Frankfurt Am Main, Germany, June 24 – July 2, 2021, Revised Selected Papers},
pages = {161–173},
numpages = {13}
}

@INPROCEEDINGS{mckenna2016ml,
  author={McKenna, Ryan and Herbein, Stephen and Moody, Adam and Gamblin, Todd and Taufer, Michela},
  booktitle={2016 IEEE International Conference on Cluster Computing (CLUSTER)}, 
  title={Machine Learning Predictions of Runtime and IO Traffic on High-End Clusters}, 
  year={2016},
  volume={},
  number={},
  pages={255-258},
  doi={10.1109/CLUSTER.2016.58}}

@INPROCEEDINGS{original_darshan_paper,
  author={Carns, Philip and Latham, Robert and Ross, Robert and Iskra, Kamil and Lang, Samuel and Riley, Katherine},
  booktitle={2009 IEEE International Conference on Cluster Computing and Workshops}, 
  title={24/7 Characterization of petascale I/O workloads}, 
  year={2009},
  volume={},
  number={},
  pages={1-10},
  doi={10.1109/CLUSTR.2009.5289150}}

@INPROCEEDINGS{darshan,
  author={Snyder, Shane and Carns, Philip and Harms, Kevin and Ross, Robert and Lockwood, Glenn K. and Wright, Nicholas J.},
  booktitle={2016 5th Workshop on Extreme-Scale Programming Tools (ESPT)}, 
  title={Modular HPC I/O Characterization with Darshan}, 
  year={2016},
  volume={},
  number={},
  pages={9-17},
  doi={10.1109/ESPT.2016.006}}

@inproceedings{daradkeh2024hpc,
author = {Daradkeh, Tariq and Roper, Gillian and Alarcon Meza, Carlos and A. Mokhov, Serguei},
title = {HPC Jobs Classification and Resource Prediction to Minimize Job Failures},
year = {2024},
isbn = {9798400716843},
publisher = {Association for Computing Machinery},
address = {New York, NY, USA},
url = {https://doi.org/10.1145/3674912.3674914},
doi = {10.1145/3674912.3674914},
booktitle = {Proceedings of the International Conference on Computer Systems and Technologies 2024},
pages = {95–101},
numpages = {7},
location = {Ruse, Bulgaria},
series = {CompSysTech '24}
}

@INPROCEEDINGS{A_Deep_Look_into_the_Temporal_IO_Behavior,
  author={Boito, Francieli and Teylo, Luan and Popov, Mihail and Jolivel, Theo and Tessier, François and Luettgau, Jakob and Monniot, Julien and Tarraf, Ahmad and Carneiro, André and Osthoff, Carla},
  booktitle={2025 IEEE International Parallel and Distributed Processing Symposium (IPDPS)}, 
  title={A Deep Look into the Temporal I/O Behavior of HPC Applications}, 
  year={2025},
  volume={},
  number={},
  pages={749-762},
  doi={10.1109/IPDPS64566.2025.00072}}

@Article{tsfresh_used_for_hydro_clustering,
AUTHOR = {Prakaisak, Intouch and Wongchaisuwat, Papis},
TITLE = {Hydrological Time Series Clustering: A Case Study of Telemetry Stations in Thailand},
JOURNAL = {Water},
VOLUME = {14},
YEAR = {2022},
NUMBER = {13},
ARTICLE-NUMBER = {2095},
URL = {https://www.mdpi.com/2073-4441/14/13/2095},
ISSN = {2073-4441},
DOI = {10.3390/w14132095}
}

@article{tsfresh_used_for_geothermal_labelling,
title = {Feature-based time series clustering for efficient labelling of geothermal data},
journal = {Geothermics},
volume = {138},
pages = {103656},
year = {2026},
issn = {0375-6505},
doi = {https://doi.org/10.1016/j.geothermics.2026.103656},
url = {https://www.sciencedirect.com/science/article/pii/S0375650526000611},
author = {Paul Michael B. Abrasaldo and Sadiq J. Zarrouk and Andreas W. Kempa-Liehr}
}

@article{tsfresh_founding_paper,
title = {Time Series FeatuRe Extraction on basis of Scalable Hypothesis tests (tsfresh – A Python package)},
journal = {Neurocomputing},
volume = {307},
pages = {72-77},
year = {2018},
issn = {0925-2312},
doi = {https://doi.org/10.1016/j.neucom.2018.03.067},
url = {https://www.sciencedirect.com/science/article/pii/S0925231218304843},
author = {Maximilian Christ and Nils Braun and Julius Neuffer and Andreas W. Kempa-Liehr}
}

@INPROCEEDINGS{recorder,
  author={Wang, Chen and Sun, Jinghan and Snir, Marc and Mohror, Kathryn and Gonsiorowski, Elsa},
  booktitle={2020 IEEE International Parallel and Distributed Processing Symposium Workshops (IPDPSW)}, 
  title={Recorder 2.0: Efficient Parallel I/O Tracing and Analysis}, 
  year={2020},
  volume={},
  number={},
  pages={1-8},
  doi={10.1109/IPDPSW50202.2020.00176}}

@article{ts_clustering_survey,
title = {Time-series clustering – A decade review},
journal = {Information Systems},
volume = {53},
pages = {16-38},
year = {2015},
issn = {0306-4379},
doi = {https://doi.org/10.1016/j.is.2015.04.007},
url = {https://www.sciencedirect.com/science/article/pii/S0306437915000733},
author = {Saeed Aghabozorgi and Ali {Seyed Shirkhorshidi} and Teh {Ying Wah}}
}

@misc{tsfresh,
  howpublished = {\url{https://tsfresh.readthedocs.io/en/latest/api/modules.html}},
  title        = {tsfresh},
  year         = {Accessed on 2026-09-12}
}

@misc{EfficientFCParameters,
  howpublished = {\url{https://tsfresh.readthedocs.io/en/latest/api/tsfresh.feature_extraction.html#tsfresh.feature_extraction.settings.EfficientFCParameters}},
  title        = {EfficientFCParameters},
  year         = {Accessed on 2026-09-12}
}

@misc{kshape,
  howpublished = {\url{https://tslearn.readthedocs.io/en/latest/gen_modules/clustering/tslearn.clustering.KShape.html}},
  title        = {KShape},
  year         = {Accessed on 2026-07-22}
}

@misc{dtw,
  howpublished = {\url{https://tslearn.readthedocs.io/en/stable/user_guide/dtw.html}},
  title        = {DTW},
  year         = {Accessed on 2026-09-11}
}

@misc{sbd,
  howpublished = {\url{https://www.aeon-toolkit.org/en/stable/api_reference/auto_generated/aeon.distances.sbd_pairwise_distance.html}},
  title        = {sbd pairwise distance},
  year         = {Accessed on 2026-07-22}
}

@misc{ts_datasets,
  howpublished = {\url{https://www.timeseriesclassification.com/dataset.php}},
  title        = {Time Series Classification Datasets},
  year         = {Accessed on 2026-09-17}
}

@ARTICLE{sleep_db_source,
  author={Kemp, B. and Zwinderman, A.H. and Tuk, B. and Kamphuisen, H.A.C. and Oberye, J.J.L.},
  journal={IEEE Transactions on Biomedical Engineering}, 
  title={Analysis of a sleep-dependent neuronal feedback loop: the slow-wave microcontinuity of the EEG}, 
  year={2000},
  volume={47},
  number={9},
  pages={1185-1194},
  doi={10.1109/10.867928}}

@article{sleep_db_site_source, title = {{PhysioNet} as a global platform for biomedical research}, volume = {1}, issn = {3005-0693}, url = {https://doi.org/10.1038/s44360-026-00096-z}, doi = {10.1038/s44360-026-00096-z}, number = {8}, journal = {Nature Health}, author = {Pollard, Tom and Moody, Benjamin E. and Lehman, Li-wei H. and Gow, Brian J. and Fernandes, Chrystinne and Xie, Chen and Johnson, Alistair and Mark, Roger G. and Heldt, Thomas}, month = aug, year = {2026}, pages = {792--795},}

@misc{pedestrian_db_source,
  howpublished = {\url{http://www.pedestrian.melbourne.vic.gov.au/#date=11-06-2018&time=4}},
  title        = {City of Melbourne Pedestrian Counting System},
  year         = {Accessed on 2026-09-17}
}

@misc{earthquakes_db_source,
  howpublished = {\url{https://www.timeseriesclassification.com/description.php?Dataset=Earthquakes}},
  title        = {Dataset: Earthquakes},
  year         = {Accessed on 2026-09-17}
}

@misc{heartbeats_db_source, author = "Bentley, P. and Nordehn, G. and Coimbra, M. and Mannor, S.", title = "The {PASCAL} {C}lassifying {H}eart {S}ounds {C}hallenge 2011 {(CHSC2011)} {R}esults", howpublished = "http://www.peterjbentley.com/heartchallenge/index.html"}

@misc{kneed,
  howpublished = {\url{https://pypi.org/project/kneed/}},
  title        = {Kneed},
  year         = {Accessed on 2026-09-08}
}

@inproceedings{boito:hal-05662894,
  TITLE = {{TOTO: Transparent I/O Tuning for HPC Applications}},
  AUTHOR = {Boito, Francieli and Teylo, Luan and Popov, Mihail and Aimi, Laora and Bandet, Alexis and Lima Pilla, La{\'e}rcio and Pallez, Guillaume},
  URL = {https://inria.hal.science/hal-05662894},
  BOOKTITLE = {{ICS 2026 - International Conference on Supercomputing}},
  ADDRESS = {Belfast, United Kingdom},
  YEAR = {2026},
  MONTH = Jul,
  DOI = {10.1145/3797905.3807867},
  HAL_ID = {hal-05662894},
  HAL_VERSION = {v1},
}

@INPROCEEDINGS{kneedle,
  author={Satopaa, Ville and Albrecht, Jeannie and Irwin, David and Raghavan, Barath},
  booktitle={2011 31st International Conference on Distributed Computing Systems Workshops}, 
  title={Finding a "Kneedle" in a Haystack: Detecting Knee Points in System Behavior}, 
  year={2011},
  volume={},
  number={},
  pages={166-171},
  doi={10.1109/ICDCSW.2011.20}}

@article{DTW_used_for_clustering,
title = {Fuzzy clustering of time series data using dynamic time warping distance},
journal = {Engineering Applications of Artificial Intelligence},
volume = {39},
pages = {235-244},
year = {2015},
issn = {0952-1976},
doi = {https://doi.org/10.1016/j.engappai.2014.12.015},
url = {https://www.sciencedirect.com/science/article/pii/S0952197614003078},
author = {Hesam Izakian and Witold Pedrycz and Iqbal Jamal}
}

@inbook{DDTW,
author = {Eamonn J. Keogh and Michael J. Pazzani},
title = {Derivative Dynamic Time Warping},
booktitle = {Proceedings of the 2001 SIAM International Conference on Data Mining (SDM)},
chapter = {},
pages = {1-11},
doi = {10.1137/1.9781611972719.1},
URL = {https://epubs.siam.org/doi/abs/10.1137/1.9781611972719.1},
eprint = {https://epubs.siam.org/doi/pdf/10.1137/1.9781611972719.1}
}

@article{betke2021classifying,
  author  = {Betke, Eugen and Kunkel, Julian},
  title   = {Classifying Temporal Characteristics of Job {I/O} Using Machine Learning Techniques},
  journal = {Journal of High-Performance Storage},
  year    = {2021}
}

@article{kim2023design,
  author    = {Kim, Sunggon and Sim, Alex and Wu, Kesheng and Byna, Suren and Son, Yongseok},
  title     = {Design and Implementation of {I/O} Performance Prediction Scheme on {HPC} Systems Through Large-Scale Log Analysis},
  journal   = {Journal of Big Data},
  volume    = {10},
  number    = {1},
  pages     = {65},
  year      = {2023},
  publisher = {Springer},
  doi       = {10.1186/s40537-023-00741-4}
}

@article{simgrid,
  author  = {Casanova, Henri and Giersch, Arnaud and Legrand, Arnaud and Quinson, Martin and Suter, Fr{\'e}d{\'e}ric},
  title   = {Versatile, Scalable, and Accurate Simulation of Distributed Applications and Platforms},
  journal = {Journal of Parallel and Distributed Computing},
  volume  = {74},
  number  = {10},
  pages   = {2899--2917},
  year    = {2014}
}

@INPROCEEDINGS{antici2026perf,
  author={Antici, Francesco and Proia, Andrea and Ohara, Ryoma and Hanawa, Toshihiro and Kiziltan, Zeynep and Bartolini, Andrea and Domke, Jens},
  booktitle={2026 IEEE 26th International Symposium on Cluster, Cloud and Internet Computing (CCGrid)}, 
  title={Automated Configuration of Power-Management Knobs for Optimal HPC Job Executions}, 
  year={2026},
  volume={},
  number={},
  pages={210-220},
  doi={10.1109/CCGrid68966.2026.00030}}

@misc{beegfs,
  title        = {{BeeGFS}: Architecture and Overview},
  author       = {{ThinkParQ GmbH}},
  howpublished = {\url{https://www.beegfs.io}},
  year         = {2018}
}

@techreport{lustre,
  author      = {Braam, Peter J. and Schwan, Philip},
  title       = {The Lustre Storage Architecture},
  institution = {Cluster File Systems, Inc.},
  year        = {2004}
}

@INPROCEEDINGS{9229642,
  author={Pottier, Loïc and da Silva, Rafael Ferreira and Casanova, Henri and Deelman, Ewa},
  booktitle={2020 IEEE International Conference on Cluster Computing (CLUSTER)}, 
  title={Modeling the Performance of Scientific Workflow Executions on HPC Platforms with Burst Buffers}, 
  year={2020},
  volume={},
  number={},
  pages={92-103},
  doi={10.1109/CLUSTER49012.2020.00019}}

@inproceedings {227665,
author = {Bin Yang and Xu Ji and Xiaosong Ma and Xiyang Wang and Tianyu Zhang and Xiupeng Zhu and Nosayba El-Sayed and Haidong Lan and Yibo Yang and Jidong Zhai and Weiguo Liu and Wei Xue},
title = {End-to-end {I/O} Monitoring on a Leading Supercomputer},
booktitle = {16th USENIX Symposium on Networked Systems Design and Implementation (NSDI 19)},
year = {2019},
isbn = {978-1-931971-49-2},
address = {Boston, MA},
pages = {379--394},
url = {https://www.usenix.org/conference/nsdi19/presentation/yang},
publisher = {USENIX Association},
month = feb
}

@INPROCEEDINGS{9139855,  
	author={M. R. {Wyatt} and S. {Herbein} and K. {Shoga} and T. {Gamblin} and M. {Taufer}},  
	booktitle={2020 IEEE International Parallel and Distributed Processing Symposium (IPDPS)},   
	title={CanarIO: Sounding the Alarm on IO-Related Performance Degradation},   
	year={2020},  
	volume={},  
	number={},  
	pages={73-83}, 
  doi={10.1109/IPDPS47924.2020.00018}}

@INPROCEEDINGS{7516071,
  author={Yildiz, Orcun and Dorier, Matthieu and Ibrahim, Shadi and Ross, Rob and Antoniu, Gabriel},
  booktitle={2016 IEEE International Parallel and Distributed Processing Symposium (IPDPS)}, 
  title={On the Root Causes of Cross-Application I/O Interference in HPC Storage Systems}, 
  year={2016},
  volume={},
  number={},
  pages={750-759},
  doi={10.1109/IPDPS.2016.50},
  ISSN={1530-2075},
  month={May},}

@article{Trochon2026OnTI,
  title={On the Impact of Interference from Concurrent Jobs on Checkpointing Performance},
  author={M{\'e}line Trochon and Jean-Thomas Acquaviva and Francieli Zanon Boito and Brice Goglin and François Tessier and Luan Teylo},
  journal={Proceedings of the 38th International Conference on Scalable Scientific Data Management},
  year={2026},
  url={https://api.semanticscholar.org/CorpusID:282922987}
}

@article{iosets,
author = {Boito, Francieli and Pallez, Guillaume and Teylo, Luan and Vidal, Nicolas},
title = {IO-Sets: Simple and Efficient Approaches for I/O Bandwidth Management},
year = {2023},
issue_date = {Oct. 2023},
publisher = {IEEE Press},
volume = {34},
number = {10},
issn = {1045-9219},
url = {https://doi.org/10.1109/TPDS.2023.3305028},
doi = {10.1109/TPDS.2023.3305028},
journal = {IEEE Trans. Parallel Distrib. Syst.},
month = oct,
pages = {2783–2796},
numpages = {14}
}

@INPROCEEDINGS{9355272,
  author={Isakov, Mihailo and Rosario, Eliakin del and Madireddy, Sandeep and Balaprakash, Prasanna and Carns, Philip and Ross, Robert B. and Kinsy, Michel A.},
  booktitle={SC20: International Conference for High Performance Computing, Networking, Storage and Analysis}, 
  title={HPC I/O Throughput Bottleneck Analysis with Explainable Local Models}, 
  year={2020},
  volume={},
  number={},
  pages={1-13},
  doi={10.1109/SC41405.2020.00037}}

@INPROCEEDINGS{8752753,
  author={Wang, Teng and Byna, Suren and Lockwood, Glenn K. and Snyder, Shane and Carns, Philip and Kim, Sunggon and Wright, Nicholas J.},
  booktitle={2019 19th IEEE/ACM International Symposium on Cluster, Cloud and Grid Computing (CCGRID)}, 
  title={A Zoom-in Analysis of I/O Logs to Detect Root Causes of I/O Performance Bottlenecks}, 
  year={2019},
  volume={},
  number={},
  pages={102-111},
  doi={10.1109/CCGRID.2019.00021}}

@inproceedings{10.1145/3581784.3607041,
author = {Karrels, Ed and Huang, Lei and Kan, Yuhong and Arora, Ishank and Wang, Yinzhi and Katz, Daniel S. and Gropp, William and Zhang, Zhao},
title = {Fine-grained Policy-driven I/O Sharing for Burst Buffers},
year = {2023},
isbn = {9798400701092},
publisher = {Association for Computing Machinery},
address = {New York, NY, USA},
url = {https://doi.org/10.1145/3581784.3607041},
doi = {10.1145/3581784.3607041},
booktitle = {Proceedings of the International Conference for High Performance Computing, Networking, Storage and Analysis},
articleno = {95},
numpages = {12},
location = {Denver, CO, USA},
series = {SC '23}
}

@inproceedings{10.1109/IPDPS.2015.116,
author = {Gainaru, Ana and Aupy, Guillaume and Benoit, Anne and Cappello, Franck and Robert, Yves and Snir, Marc},
title = {Scheduling the I/O of HPC Applications Under Congestion},
year = {2015},
isbn = {9781479986491},
publisher = {IEEE Computer Society},
address = {USA},
url = {https://doi.org/10.1109/IPDPS.2015.116},
doi = {10.1109/IPDPS.2015.116},
booktitle = {Proceedings of the 2015 IEEE International Parallel and Distributed Processing Symposium},
pages = {1013–1022},
numpages = {10},
series = {IPDPS '15}
}

@INPROCEEDINGS{7307592,
  author={Zhou, Zhou and Yang, Xu and Zhao, Dongfang and Rich, Paul and Tang, Wei and Wang, Jia and Lan, Zhiling},
  booktitle={2015 IEEE International Conference on Cluster Computing}, 
  title={I/O-Aware Batch Scheduling for Petascale Computing Systems}, 
  year={2015},
  volume={},
  number={},
  pages={254-263},
  doi={10.1109/CLUSTER.2015.45}}

@INPROCEEDINGS{7877148,
  author={Liu, Yang and Gunasekaran, Raghul and Ma, Xiaosong and Vazhkudai, Sudharshan S.},
  booktitle={SC '16: Proceedings of the International Conference for High Performance Computing, Networking, Storage and Analysis}, 
  title={Server-Side Log Data Analytics for I/O Workload Characterization and Coordination on Large Shared Storage Systems}, 
  year={2016},
  volume={},
  number={},
  pages={819-829},
  doi={10.1109/SC.2016.69}}

@INPROCEEDINGS{10579160,
  author={Tarraf, Ahmad and Bandet, Alexis and Boito, Francieli and Pallez, Guillaume and Wolf, Felix},
  booktitle={2024 IEEE International Parallel and Distributed Processing Symposium (IPDPS)}, 
  title={Capturing Periodic I/O Using Frequency Techniques}, 
  year={2024},
  volume={},
  number={},
  pages={465-478},
  doi={10.1109/IPDPS57955.2024.00048}}

@INPROCEEDINGS{9820667,
  author={Yang, Bin and Zou, Yanliang and Liu, Weiguo and Xue, Wei},
  booktitle={2022 IEEE International Parallel and Distributed Processing Symposium (IPDPS)}, 
  title={An End-to-end and Adaptive I/O Optimization Tool for Modern HPC Storage Systems}, 
  year={2022},
  volume={},
  number={},
  pages={1294-1304},
  doi={10.1109/IPDPS53621.2022.00128}}

@INPROCEEDINGS{9820616,
  author={Yang, Wenxiang and Liao, Xiangke and Dong, Dezun and Yu, Jie},
  booktitle={2022 IEEE International Parallel and Distributed Processing Symposium (IPDPS)}, 
  title={A Quantitative Study of the Spatiotemporal I/O Burstiness of HPC Application}, 
  year={2022},
  volume={},
  number={},
  pages={1349-1359},
  doi={10.1109/IPDPS53621.2022.00133}}

@INPROCEEDINGS{9910076,
  author={Costa, Emily and Patel, Tirthak and Schwaller, Benjamin and Brandt, Jim M. and Tiwari, Devesh},
  booktitle={SC21: International Conference for High Performance Computing, Networking, Storage and Analysis}, 
  title={Systematically Inferring I/O Performance Variability by Examining Repetitive Job Behavior}, 
  year={2021},
  volume={},
  number={},
  pages={1-15},
  doi={10.1145/3458817.3476186}}

@ARTICLE{7289462,
  author={Dorier, Matthieu and Ibrahim, Shadi and Antoniu, Gabriel and Ross, Rob},
  journal={IEEE Transactions on Parallel and Distributed Systems}, 
  title={Using Formal Grammars to Predict I/O Behaviors in HPC: The Omnisc'IO Approach}, 
  year={2016},
  volume={27},
  number={8},
  pages={2435-2449},
  doi={10.1109/TPDS.2015.2485980}}

@INPROCEEDINGS{9460487,
  author={Bez, Jean Luca and Miranda, Alberto and Nou, Ramon and Boito, Francieli and Cortes, Toni and Navaux, Philippe},
  booktitle={2021 IEEE International Parallel and Distributed Processing Symposium (IPDPS)}, 
  title={Arbitration Policies for On-Demand User-Level {I/O} Forwarding on {HPC} Platforms}, 
  year={2021},
  volume={},
  number={},
  pages={577-586},
  doi={10.1109/IPDPS49936.2021.00066}
  }

@inproceedings{bandet:hal-04394004,
  TITLE = {{Scheduling distributed I/O resources in HPC systems}},
  AUTHOR = {Bandet, Alexis and Boito, Francieli and Pallez, Guillaume},
  URL = {https://inria.hal.science/hal-04394004},
  BOOKTITLE = {{30th International European Conference on Parallel and Distributed Computing 26 - 30 August 2024 Madrid, Spain 30th International European Conference on Parallel and Distributed Computing}},
  ADDRESS = {Madrid, Spain},
  YEAR = {2024},
  MONTH = Aug,
  HAL_ID = {hal-04394004},
  HAL_VERSION = {v2},
}

\end{document}